\documentclass[fleqn,usenatbib]{mnras}

\usepackage{newtxtext,newtxmath}

\usepackage[T1]{fontenc}

\DeclareRobustCommand{\VAN}[3]{#2}
\let\VANthebibliography\thebibliography
\def\thebibliography{\DeclareRobustCommand{\VAN}[3]{##3}\VANthebibliography}

\usepackage[pdftex]{graphicx}
\usepackage{enumitem}
\usepackage{mathtools}
\usepackage[version=4]{mhchem}
\usepackage{siunitx}
\usepackage{booktabs}
\usepackage{amsmath,bm}
\usepackage{graphicx}

\DeclarePairedDelimiterX\braket[2]{\langle}{\rangle}{#1 \delimsize\vert #2}
\usepackage{float}
\usepackage{array,multirow,graphicx}
\usepackage[dvipsnames]{xcolor}
\usepackage{euscript}
\definecolor{mygray}{gray}{0.5}

\title[Global 21-cm: Flex Your Knots]{FlexKnot as a Generalised Model of the Sky-averaged 21-cm Signal at \bm{$z \sim 6-30$} in the Presence of Systematics}

\author[E. Shen et al.]{
Emma Shen,$^{1,2}$\thanks{E-mail: yhs24@cam.ac.uk}
Dominic Anstey,$^{1,2}$
Eloy de Lera Acedo$^{1,2}$
and Anastasia Fialkov$^{2,3}$
\\
$^{1}$Cavendish Laboratory, University of Cambridge, Cambridge, CB3 0HE, United Kingdom\\
$^{2}$Kavli Institute for Cosmology, Madingley Road, Cambridge, CB3 0HA, United Kingdom\\
$^{3}$Institute of Astronomy, University of Cambridge, Madingley Road, Cambridge CB3 0HA, United Kingdom
}

\date{Accepted XXX. Received YYY; in original form ZZZ}

\pubyear{2023}

\begin{document}
\label{firstpage}
\pagerange{\pageref{firstpage}--\pageref{lastpage}}
\maketitle

\begin{abstract}
Global 21-cm experiments are built to study the evolution of the Universe between the cosmic dawn and the epoch of reionisation. FlexKnot is a function parameterised by freely moving knots stringed together by splines. Adopting the FlexKnot function as the signal model has the potential to separate the global 21-cm signal from the foregrounds and systematics while being capable of recovering the crucial features given by theoretical predictions. In this paper, we implement the FlexKnot method by integrating twice over a function of freely moving knots interpolated linearly. The function is also constrained at the lower frequencies corresponding to the dark ages by theoretical values. The FlexKnot model is tested in the framework of the realistic data analysis pipeline of the REACH global signal experiment using simulated antenna temperature data. We demonstrate that the FlexKnot model performs better than existing signal models, e.g. the Gaussian signal model, at reconstructing the shape of the true signals present in the simulated REACH data, especially for injected signals with complex structures. The capabilities of the FlexKnot signal model is also tested by introducing various systematics and simulated global signals of different types. These tests show that four to five knots are sufficient to recover the general shape of most realistic injected signals, with or without sinusoidal systematics. We show that true signals whose absorption trough of amplitude between 120 to 450 mK can be well recovered with systematics up to about 50 mK.

\end{abstract}

\begin{keywords}
methods: data analysis -- dark ages, reionisation, first stars
\end{keywords}



\section{Introduction}

The cosmic epoch between recombination and reionisation is still a missing piece in modern cosmology. This period of cosmic history sees the Universe through the dark ages, the first light, and then the onset of cosmic reionisation. The hyperfine transition of neutral hydrogen of wavelength $\lambda = 21$ cm in its rest frame is an observable that can trace the local properties of gas in the intergalactic medium over time. A change in the brightness temperature relative to the radio background can occur when neutral hydrogen absorbs or emits at this wavelength, the level of which varies with redshift.

Existing observatories in an attempt to detect the 21-cm hydrogen line are designed following two main approaches. First, the interferometric instruments such as the SKA \citep{dd2009}, HERA \citep{hera2017}, PAPER \citep{parsons2010}, MWA \citep{mwa2009}, LOFAR \citep{lofar}, and NeuFAR \citep{neufar} are designed to detect full spatially varying power spectrum of the 21-cm hydrogen line. The second, conceptually simpler but harder to calibrate approach adopts a wide-beam single antenna system to detect the spatially averaged, or sky-averaged, global 21-cm signal. These experiments include EDGES \citep{edges}, SARAS \citep{saras, saras3},  BIGHORNS \citep{bhorn}, MIST \citep{mist}, and REACH \citep{reachh, eloynat}.

The hyperfine transition of neutral hydrogen can be considered in terms of spin temperature, $T_\mathrm{S}$, which describes the relative populations of the two energy levels. It can be defined as:
\begin{equation}
\begin{aligned}
\frac{n_1}{n_0} = \frac{g_1}{g_0} e^{-\frac{T_*}{T_\mathrm{S}}},
\end{aligned}
\label{eqspin}
\end{equation}
where $n_1$ and $n_0$ are the relative populations of the higher and lower energy levels, $g_1$ and $g_0$ are the degeneracies of the higher and lower energy levels, respectively, and $T_*$ is the transition energy divided by the Boltzmann constant.

\begin{figure}
    \centering
    \minipage{0.5\textwidth} \includegraphics[trim={0cm 0cm 1.2cm 0cm},clip,width=\linewidth]{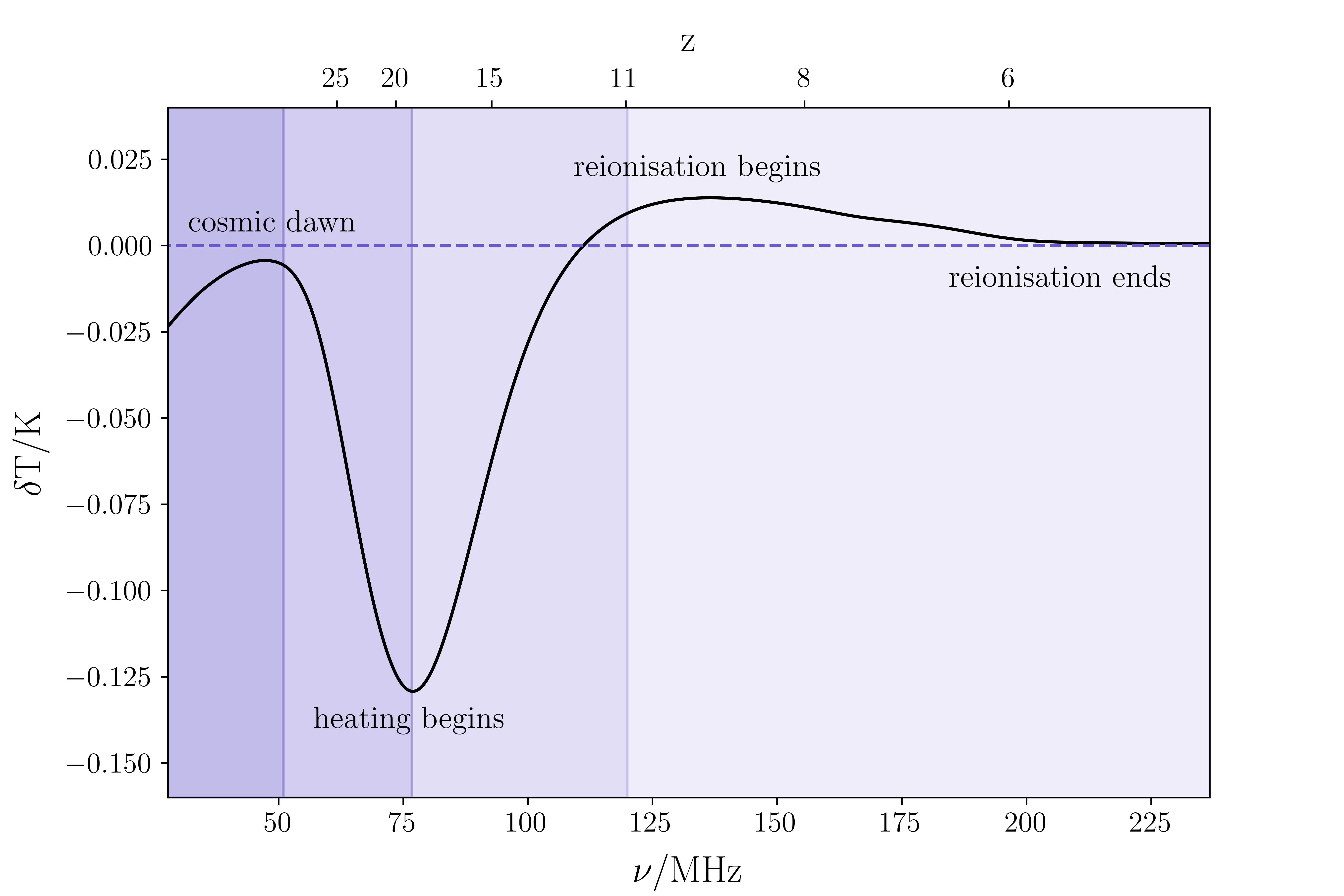}
    \endminipage\hfill
    \caption{A simulated global 21-cm signal generated using \textsc{globalemu} \citep{bevins2021} with the following parameters: star formation efficiency $f_* = 0.02$, minimal virial circular velocity $V_\mathrm{c} = 16.5$ $\mathrm{km}/\mathrm{s}$, X-ray efficiency $f_\mathrm{x}= 1$, CMB optical depth $\tau= 0.06$, slope of the spectral energy density $\alpha= 1$, low energy cut-off of the X-ray spectral energy density $\nu_\mathrm{min}= 0.2$ $\mathrm{keV}$, mean free path of the ionising photons $R_\mathrm{mfp}= 30$ $\mathrm{Mpc}$. The major cosmic events are marked on the plot, namely, cosmic dawn (the onset of star formation), heating, and the beginning and the end of reionisation.}
    \label{figsignal}
\end{figure}

The true shape of the global 21-cm signal during the epoch of reionisation (EoR) is yet to be known. An example of the simulated signal by \textsc{globalemu} \citep{bevins2021} is shown in Fig. \ref{figsignal}: during the dark ages, gas is sufficiently dense for collisions to couple the spin temperature of the hydrogen gas to the kinetic temperature of the gas, $T_\mathrm{S} \rightarrow T_\mathrm{K}$. Due to adiabatic cooling of gas, the kinetic temperature of the gas is lower than the background radiation temperature, $T_\mathrm{K} < T_{\gamma}$ which is usually assumed to be the cosmic microwave background (CMB) temperature. This difference results in an absorption against the radio background, making $T_\mathrm{S} < T_{\gamma}$. The collisional transitions become negligible as gas density decreases due to continuing expansion, setting $T_\mathrm{S} \rightarrow T_{\gamma}$. During cosmic dawn, the first luminous objects begin to emit radiation in the Lyman-band. The spin temperature is then coupled to cold gas, $T_\mathrm{S} \sim T_\mathrm{K} < T_{\gamma}$ via Wouthuysen-Field effect \citep{wouw, field}, resulting in another absorption. Fluctuations as well as total intensity in the Lyman-$\alpha$ background no longer affect the 21-cm signal after Lyman-$\alpha$ coupling saturates. The growing population of X-ray sources start heating the adiabatically cooling gas, driving $T_\mathrm{S} \sim T_\mathrm{K} > T_{\gamma}$ in general. In some models, this would turn absorption into emission. Ionising photons begin to turn neutral hydrogen (HI) to ionised hydrogen (HII), and then the signal eventually tends to zero \citep{shaver, furl}. Knowing the location and the shape of the global 21-cm signal absorption trough can enlighten us on the timing of the primordial star formation, the mass and star formation efficiency of the first star forming halos, and the luminosity of the first X-ray sources \citep{furl2006,mcquinn2012}.

The EDGES group \citep{edges} reported to have detected an absorption profile in the shape of a flattened Gaussian centred at 78 MHz with an amplitude of 500 mK in the sky-averaged 21-cm spectrum. Re-examination of the EDGES data analysis  \citep{hillsnat, singh}, however, has pointed out the non-physical parameters, the non-uniqueness of their solution, as well as  potentially unaccounted for systematic structures in the data. Studies such as \citet{sims} and \citet{bevins} have shown that a damped sinusoidal systematic is strongly preferred in the EDGES data. Further observations such as SARAS3 also reported non-detection of the EDGES profile with 95.3\% confidence \citep{singh22}.

Existing methods in data analysis and interpretation such as using the shape of a Gaussian or a flattened Gaussian as a signal model cannot describe both absorption and emission peaks at the same time. On ther other hand, physically modelled signals \citep{21cmfast} such as those of 21cmGEM \citep{21gem} (available at \url{https://people.ast.cam.ac.uk/~afialkov/Publications.html}), and \textsc{globalemu} \citep{bevins2021} have limited signal trough without enhanced radio backgrounds. In this paper, we adopt and test a signal model that has the potential of describing a variety of shapes while still having the ability to separate the signal from the foregrounds as well as the systematics. This is realised by adopting a function parameterised by freely moving knots stringed together by splines called the FlexKnot model. The FlexKnot method is introduced by \citet{vaz} to reconstruct the primordial power spectrum from the Planck data, and then adopted by \citet{millea} for parameterising the reionisation history. \citet{heimer22} used it to parameterise the EoR history and the potential of hypothetical high-redshift Fast Radio Bursts to constrain cosmic reionisation. 

Recently, \citet{heimer23} has also adopted the FlexKnot method as a global signal model. They used FlexKnot to separate the non-foreground component using EDGES low-band data. There are several differences between our implementation and the approach of \citet{heimer23}. In contrast to the work we present here, they do not distinguish between cosmological signal and systematics. They modelled the foregrounds using a polynomial, while we simulate the foregrounds with a physically motivated method. Moreover, instead of adopting Piecewise Cubic Hermite Interpolating Polynomial (\textsc{pchip}) to perform interpolation like in aforementioned works, our model starts by parameterising the second derivative of the global 21-cm signal via linearly interpolated splines, and the signal is recovered by integrating it twice.  We test the FlexKnot signal model by implementing it in the REACH data analysis pipeline \citep{anstey} where the physically motivated foreground and the signal are jointly fitted using the Bayesian nested sampling algorithm \textsc{PolyChord} \citep{hand5b}. We also test the FlexKnot model on cases in which uncalibrated sinusoidal systematic structures are present. Unaccounted for systematics would potentially arise in practice, due to e.g. ground plane artefact \citep{brad} and calibration issues \citep{sims}, so it is crucial to know how susceptible the signal model is to the presence of systematics.

This paper is organised as follows. In Section \ref{sec2}, we detail how the function is parameterised and introduce the additional constraints applied to the signal model. Section \ref{sec3} briefly describes how the foreground is modelled in the REACH data analysis pipeline as well as Bayesian inference on which the pipeline is based. The results are presented in Section \ref{sec4}, which covers comparison between different signal models, signal recovery in the presence of sinusoidal systematics in the simulated antenna temperature data, and lastly the optimal number of knots. Section \ref{seccon} concludes the work.

\section{Constructing the FlexKnot Function}
\label{sec2}
\begin{figure}
    \centering
    \minipage{0.5\textwidth} \includegraphics[trim={0.99cm 0cm 0.99cm 0cm},clip,width=\linewidth]{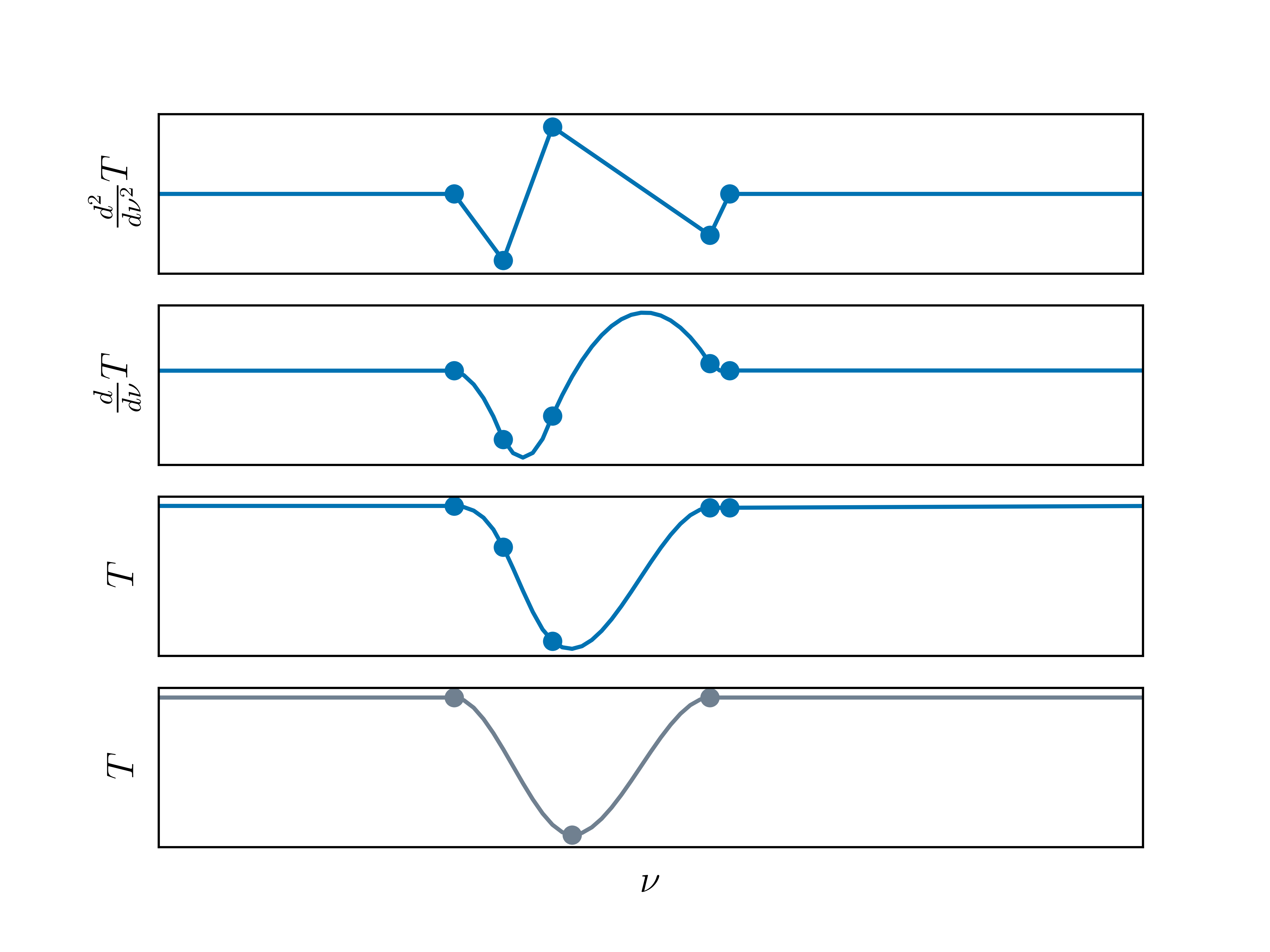}
    \endminipage\hfill
    \caption{The upper three panels (blue)  illustrate how the FlexKnot function is implemented in our signal model in three steps: the second derivative of the global 21-cm signal is parameterised by linearly interpolated splines, shown in the first panel, and the signal is recovered by integrating it twice, shown respectively in the second and the third panel. The bottom panel (gray) shows a similar signal directly fitted by Piecewise Cubic Hermite Interpolating Polynomial (\textsc{pchip}). To recover a Gaussian-like signal, the number of knots required in the two different methods is different, which can be seen in the last two panels.}
    \label{figsd}
\end{figure}

The FlexKnot model used in this paper is based on \citet{vaz}, who proposed the model to reconstruct the primordial power spectrum from the Planck data. It has also been adopted in several other works for its higher degree of flexibility \citep{bridges,vazq,vaz2013,yan2014,jian,hee,planck15,hee17,ola18,millea,planck20,planck202,heimer22,esca23}. In this paper, it is formed by a number of knots interpolated by splines. It has the advantage of describing functions of all possible shapes, given sufficient number of knots; this attribute is desirable because the exact shape of the global 21-cm signal is yet to be known. Normally, there is a deep absorption trough at $z\sim 10-30$ and an emission at $z\sim 6-10$; but this is model-dependent. To model the global 21-cm signal, piecewise cubic spline interpolation seems to be the preferable choice. In this work, the FlexKnot function is constructed by integrating twice over a function of freely moving knots interpolated by piecewise linear splines:
\begin{equation}
\begin{aligned}
f_{\mathrm{cubic}}(\nu) = \int^{\nu}_{\nu_0} d\nu'\int^{\nu'}_{\nu'_0} d\nu'' f_{\mathrm{linear}}(\nu''),
\end{aligned}
\label{eqintegration}
\end{equation}
which is the equivalent to a function made up of piecewise cubic splines, and works round the difficulties in interpreting the posteriors and setting priors when directly employing cubic spline interpolation with nested sampling \citep{Handley20189}. The positions of the knots are determined by Bayesian nested sampling. Fig. \ref{figsd} shows an example of a signal recovered by the two different methods: the upper three panels show how the FlexKnot function is implemented as a signal model in three steps, and the last panel shows the direct cubic spline interpolation done via Piecewise Cubic Hermite Interpolating Polynomial (\textsc{pchip}). To recover a signal alike to a Gaussian, the number of knots required in the two different methods is different. The reconstructed Gaussian-like signal in Fig. \ref{figsd} requires two more knots for the second derivative parameterisation method to recover a Gaussian signal with reasonable accuracy than the direct cubic spline interpolation method.

\begin{figure}
    \centering
    \minipage{0.5\textwidth} \includegraphics[trim={0.99cm 0cm 0.99cm 0cm},clip,width=\linewidth]{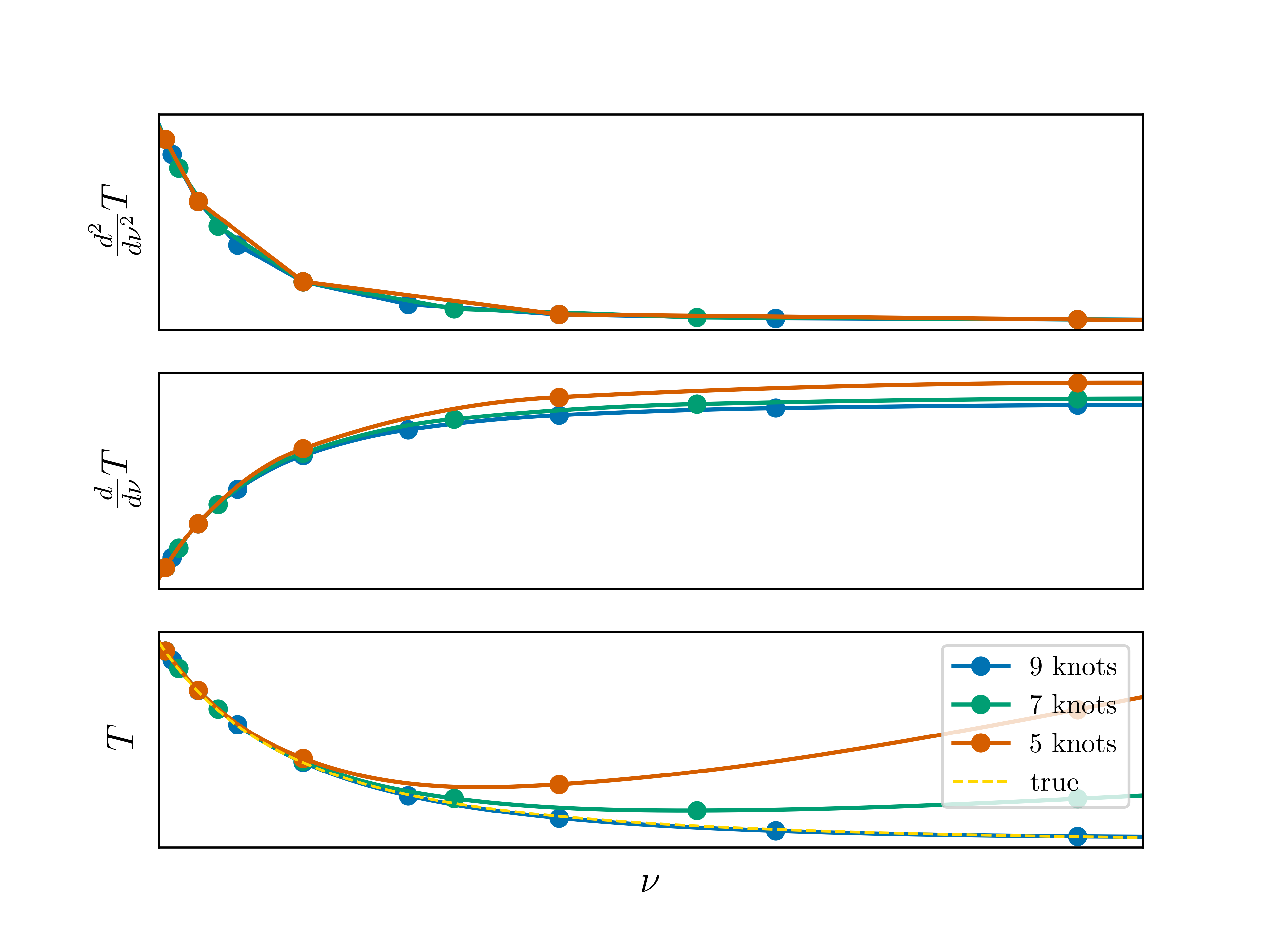}
    \endminipage\hfill
    \caption{This figure shows how the FlexKnot model of different $N_{\mathrm{knot}}$ recovers the signal in three steps in the case where it aims to recover a power law function $T(\nu) = \nu^{-\frac{5}{2}}$ represented by the yellow dashed line: the second derivative of the global 21-cm signal is parameterised by linearly interpolated splines, shown in the first panel, and the signal is recovered by integrating it twice, shown respectively in the second and the third panel. Unlike functions with only a peak or trough, a power law function requires a larger number of knots distributed throughout the argument range to be properly described.}
    \label{figsdp}
\end{figure}

The FlexKnot signal model is then implemented in the REACH data analysis pipeline to test its functionality. The REACH data analysis pipeline, based on Bayesian nested sampling, is detailed in section \ref{sec3}. In the cases of only a few knots, the FlexKnot function modelled via parameterising the second derivative is suitable for describing shapes with obvious troughs or crests, such as the shape of the global 21-cm signal, which, typically, is predicted to have one bigger trough and an ensuing emission between $z\sim 6-30$. To describe a smooth shape like the power law function accurately, it would require a much higher number of knots distributed throughout the argument range, as the second derivative of a power law is non-linear except when the order is 2 or 3. This characteristic can also potentially help separate the global 21-cm signal from the smooth foregrounds. Fig. \ref{figsdp} shows an example of how the FlexKnot model of different $N_{\mathrm{knot}}$ aims to recover a power law function:
\begin{equation}
\begin{aligned}
T(\nu) = \nu^{-\frac{5}{2}}.
\end{aligned}
\label{eqpower}
\end{equation} 
In this case, it requires 9 knots for it to be properly recovered, and a lower number of knots leads to inaccurate results. The optimal number of knots to use in each case would also depend on the step size in numerical integration.

\subsection{Parameters and Priors}
Each knot $K_{i}$ is defined by two parameters: the position in the observing frequency and the second derivative in temperature. There is an additional highest frequency knot $K_{N+1}$ fixed at 0 K after which the integrated value remains zero, as it is known that reionisation will eventually eliminate the 21-cm signal. This  additional highest frequency knot is excluded from the total number of knots $N_{\mathrm{knot}}$. The total number of parameters $n_{\mathrm{signal}}$ of the FlexKnot signal model is then:
\begin{equation}
\begin{aligned}
n_{\mathrm{signal}} = 2 \times N_{\mathrm{knot}} + 1.
\end{aligned}
\label{eqparano}
\end{equation} 

As the FlexKnot signal model is built in the mindset of the Bayesian nested sampling technique, priors need to be set for the parameters. A sorted prior is adopted for the positional parameter of the knots so that the positional parameters would always be in ascending order. The prior range of the positional parameters is [$50, 200$] MHz (sorted uniform prior), which covers the observational frequency range of the REACH experiment. 

For the second derivative parameters, the prior is set to be [$-0.02, 0.02$] (uniform prior); the prior limits should be adjusted depending on the interval between the frequency points, as the derivative describes the rate of change between two discrete points in this model. In our tests, the interval between the frequency points is set to 1 MHz. The second derivative of the lowest frequency point (the first knot $K_1$), however, is set to be negative only [$-0.02, 0$] (uniform prior) to prevent the function from turning uncharacteristically high at the low frequencies  corresponding to higher redshifts, which could potentially occur when the chromatic foregrounds in the form of power law function are present in the data. The additional highest frequency knot $K_{N+1}$ does not have the second derivation parameter; its twice integrated value is fixed at 0 K, with both its first and second derivative being also null.

\subsection{Theoretical dark ages Primer}
\begin{figure}
    \centering
    \minipage{0.49\textwidth} \includegraphics[trim={0cm 0cm 0 0cm},clip,width=\linewidth]{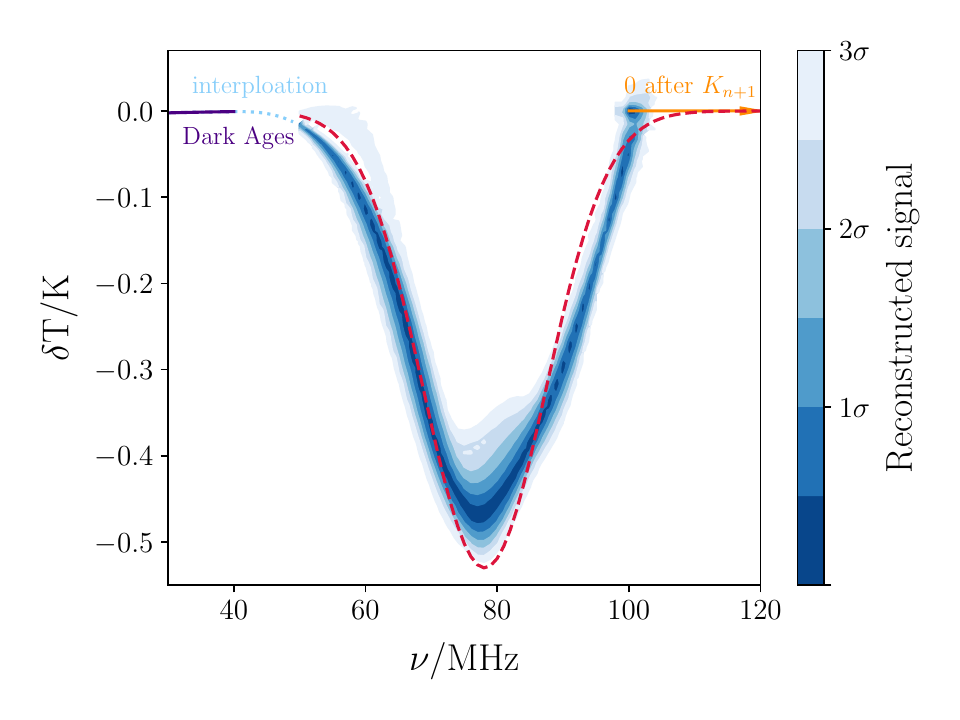}
    \endminipage\hfill
    \caption{An example of the reconstructed signal (predictive posterior of the final function given the distribution of the parameter posteriors) in blue contours by the REACH data analysis pipeline using a 4-knot FlexKnot signal model with constraints including the dark ages primer (theoretical global 21-cm signal values within the dark ages frequencies), between 30 MHz and 40 MHz (purple solid line), negative prior constraint on the first knot, and null values after the highest frequency knot $K_{N+1}$ (orange arrow). The red-dashed line represents the simulated signal in the shape of a Gaussian, and the blue-dotted line traces the interpolation between 40 MHz and 50 MHz, the end point of the dark ages primer and the beginning of the reconstructed signal.}
    \label{figflex}
\end{figure}

\begin{figure}
    \centering
    \minipage{0.49\textwidth} \includegraphics[trim={0cm 0cm 0 0cm},clip,width=\linewidth]{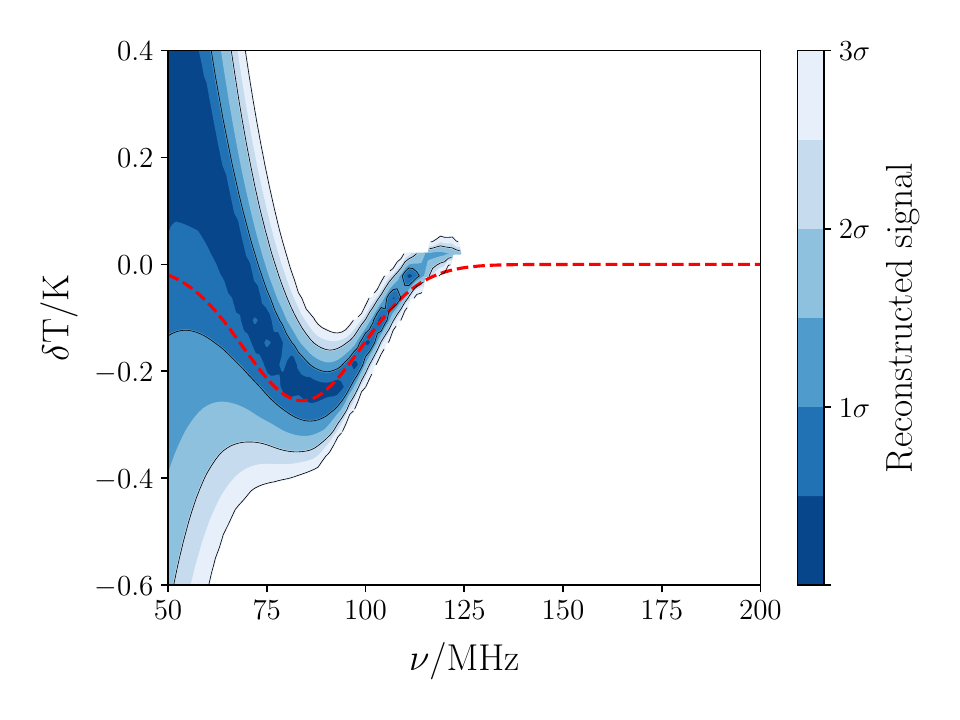}
    \endminipage\hfill
    \caption{An example of the reconstructed signal (predictive posterior of the final function given the distribution of the parameter posteriors) using a 6-knot FlexKnot signal model with neither prior constraint on the first knot nor the theoretical dark ages primer (theoretical global 21-cm signal values within the dark ages frequencies). The red-dashed line represents the simulated injected signal. The reconstructed signal in blue contours diverges at the lower frequency end, otherwise being reasonably well recovered, such as the centre frequency, absorption level, and width.}
    \label{figbflex}
\end{figure}

Unlike during the EoR, the theoretical global 21-cm signal is better understood during the dark ages \citep{hogan1979,scott1990,mondal2023}. To take advantage of that, in our model (Fig. \ref{figflex}), a set of  theoretical values at frequencies corresponding to the dark ages ($z \sim 30 - 40 $) lying outside the observational frequency range of REACH is implemented as a primer, or the first segment that initialises the ensuing function, to not only provide a reasonable and reliable constraint, but also further prevent potential uncharacteristic positive surge at the lower frequencies that sometimes occurs, as the case shown in Fig. \ref{figbflex}. The low frequency end of the function is then interpolated between the last point of the theoretical dark ages primer and the first knot. The value of the theoretical dark ages signal is generated by the \textsc{globalemu} emulator \citep{bevins2021} modified for the dark ages where different parameters are used. The global 21-cm signal during the dark ages is characterised by the following astrophysical parameters: baryon density, $\Omega_b = 0.01495074$, matter density, $\Omega_m = 0.29395689$, curvature density, $\Omega_k = 0.05966342$, and reduced Planck value, $h = 0.75665809 \: \mathrm{km}\: \mathrm{s}^{-1}\mathrm{Mpc}^{-1}$ .

\section{Data Analysis Pipeline}
\label{sec3}
To test the FlexKnot signal model, the REACH data analysis pipeline \citep{anstey} is adopted to extract the injected global 21-cm signal from the simulated antenna temperature data (section \ref{sec32}). Simulated data that include the foregrounds, the injected global signal, systematics, and noise would be generated and then fed into the pipeline for it to separate the signal from other elements. The pipeline is built with the purpose to avoid degeneracy with potential systematics by adopting physically motivated foreground modelling, where the foreground model and the instrument model are jointly fitted with the signal model using the Bayesian inference technique. The Bayesian nested sampling algorithm \textsc{PolyChord} is opted for its ability to estimate Bayesian evidence efficiently in the case of high model dimensionalities, which is inherent to this implementation. 

\subsection{Bayesian Inference}
Bayesian inference is a statistical modelling technique that has its merit in parameter estimation and model comparison. A model $\mathcal{M}$ parameterised by $\theta_\mathcal{M}$ can be used to calculate the probability of observing the data $\mathcal{D}$ by updating previous knowledge of the parameters, the prior $\mathrm{P}(\theta_\mathcal{M}|\mathcal{M})$. This can be done by applying Bayes' theorem,
\begin{equation}
\begin{aligned}
\mathrm{P}(\theta_\mathcal{M}|\mathcal{D},\mathcal{M}) &= \frac{\mathrm{P}(\mathcal{D}|\theta_\mathcal{M},\mathcal{M})\mathrm{P}(\theta_\mathcal{M}|\mathcal{M})}{\mathrm{P}(\mathcal{D}|\mathcal{M})}, \quad \mathrm{or} \quad
\mathcal{P} = \frac{\mathcal{L}\pi}{\mathcal{Z}},
\end{aligned}
\label{eqbayes}
\end{equation}
where $\mathcal{P}$ is the posterior distribution, $\mathcal{L}$ is the likelihood, the probability of the data given a model and the set of parameters describing the model, $\mathcal{\pi}$ is the prior distribution of the parameters, and $\mathcal{Z}$ is the Bayesian evidence or marginal likelihood, which gives the probability of observing the data $\mathcal{D}$ given the model $\mathcal{M}$ \citep{Sivia2006}. One can achieve marginalisation by integrating over the prior distribution:
\begin{equation}
\begin{aligned}
\mathcal{Z} = \int \mathrm{P}(\mathcal{D}|\theta_\mathcal{M},\mathcal{M})\mathrm{P}(\theta_\mathcal{M}|\mathcal{M})d\theta_\mathcal{M} = \int \mathcal{L}\pi d\theta_\mathcal{M}.
\end{aligned}
\label{eqev}
\end{equation}
To compare different models, one can derive the probability of a model given the data by applying Bayes' theorem on the Bayesian evidence:
\begin{equation}
\begin{aligned}
\mathrm{P}(\mathcal{M}|\mathcal{D}) = \frac{\mathrm{P}(\mathcal{D}|\mathcal{M})\mathrm{P}(\mathcal{M})}{\mathrm{P}(\mathcal{D})} = \mathcal{Z} \frac{\mathrm{P}(\mathcal{M})}{\mathrm{P}(\mathcal{D})},
\end{aligned}
\label{eqmodel}
\end{equation}
where $\mathrm{P}(\mathcal{D})$ is a normalisation factor independent of the model. As such, one may compare two competing models $\mathcal{M}_1$ and $\mathcal{M}_2$ by taking the ratio of the two evidences weighted by $\mathrm{P}(\mathcal{M})$, or by taking the logarithmic Bayes factor:
\begin{equation}
\begin{aligned}
\Delta \log(\mathcal{Z}) = \log\mathrm{P}(\mathcal{M}_2|\mathcal{D}) - \log\mathrm{P}(\mathcal{M}_1|\mathcal{D}),
\end{aligned}
\label{eqmodelprefer}
\end{equation}
under the assumption of uniform weighting $\mathrm{P}(\mathcal{M}_1) = \mathrm{P}(\mathcal{M}_2)$. A positive $\Delta \log(\mathcal{Z})$ indicates the preference of model $\mathcal{M}_2$ over $\mathcal{M}_1$ with betting odds of $e^{\Delta \log(\mathcal{Z})}:1$. In the context of this work, we would first take the $\Delta \log(\mathcal{Z})$ between the model with a signal:
\begin{equation}
\begin{aligned}
\mathcal{M}_n = \mathcal{M}_{\mathrm{fg}} + \mathcal{M}_{\mathrm{21}} + \mathcal{M}_{\mathrm{noise}},
\end{aligned}
\label{eqmn}
\end{equation}
and the one without:
\begin{equation}
\begin{aligned}
\mathcal{M}_0 = \mathcal{M}_{\mathrm{fg}} + \mathcal{M}_{\mathrm{noise}},
\end{aligned}
\label{eqm0}
\end{equation} to first make sure the model with a signal is indeed statistically favourable before proceeding to compare the $\Delta \log(\mathcal{Z})$ yielded by the different signal models. $\Delta \log(\mathcal{Z})$ is also referred to as the log evidence in this work.

\subsection{Simulated Antenna Temperature Data}
\label{sec32}
The simulated antenna temperature data generated to test the capability of the FlexKnot signal model to extract the global 21-cm signal in the REACH data analysis pipeline include the following four components: the foregrounds, the sky-averaged 21-cm signal, the systematics, and the antenna temperature noise:
\begin{equation}
\begin{aligned}
T_{\mathrm{data}} = T_{\mathrm{fg}} + T_{21} + T_{\mathrm{sys}} + T_\mathrm{noise}.
\end{aligned}
\label{eqantda}
\end{equation}

\subsubsection{Foreground}
\label{sec321}
The spatially varying spectral index map is simulated by the sky model based on instances of the 2008 Global Sky Model \citep{gsm2008} at 408 MHz and 230 MHz:
\begin{equation}
\begin{aligned}
\beta (\theta,\phi) = \frac{\log\left( \frac{T_\mathrm{230}(\theta,\phi) - T_\mathrm{CMB}}{T_\mathrm{408}(\theta,\phi) - T_\mathrm{CMB}} \right)}{\log \left( \frac{230}{480}\right)},
\end{aligned}
\label{eqspec}
\end{equation}
with which the sky model can be generated:
\begin{equation}
\begin{aligned}
T_\mathrm{sky} (\theta,\phi,\nu) = \left( T_\mathrm{480} (\theta,\phi) - T_\mathrm{CMB} \right) \left(\frac{\nu}{408}\right)^{-\beta (\theta,\phi)} + T_\mathrm{CMB}.
\end{aligned}
\label{eqsky}
\end{equation}
It is then convolved with the beam pattern of a conical log spiral antenna \citep{dyson1965} of beam pattern $D(\theta,\phi,\nu)$ to generate the foreground component: 
\begin{equation}
\begin{aligned}
T_\mathrm{fg} = \int_{\Omega} D(\theta,\phi,\nu) T_\mathrm{sky} (\theta,\phi,\nu) d\Omega.
\end{aligned}
\label{eqfore}
\end{equation}

\subsubsection{Sky-averaged 21-cm Signal}
\label{secsignal}
The Gaussian signal, the flattened Gaussian signal, and the signals emulated by \textsc{globalemu} are the three different types of signal that have been injected as the true signal in the simulated antenna temperature data in this work.

A Gaussian is generally written as 
\begin{equation}
\begin{aligned}
T_{21,\mathrm{g}}(\nu) = -A_{21}\left( e^{\frac{-(\nu - \nu_0)^ 2}{2 \sigma^2}}\right),
\end{aligned}
\label{eqgau}
\end{equation}
where $A_{21}$ is the absorption amplitude, $\nu_0$ is the centre frequency, and $\sigma$ is the standard deviation. The explored parameter space of the injected signal in the form of Gaussian is as follows:
\begin{itemize}
    \item [] \textbullet\hspace{\labelsep} \makebox[3cm][l]{amplitude:} \makebox[2.cm][r]{\{155, 255, 455\}} \: mK
    \item [] \textbullet\hspace{\labelsep} \makebox[3cm][l]{centre frequency:} \makebox[2.cm][r]{\{85, 105, 125\}} \: MHz
    \item [] \textbullet\hspace{\labelsep} \makebox[3cm][l]{standard deviation:} \makebox[2.cm][r]{\{10, 15, 20\}} \: MHz
\end{itemize}

\citet {edges} defined the flattened Gaussian function by the equation:
\begin{equation}
\begin{aligned}
T_{21,\mathrm{fg}}(\nu) = -A_{21}\left(\frac{1-e^{-\tau e^{B}}}{1-e^{-\tau}}\right),
\end{aligned}
\label{eqfga}
\end{equation}
with
\begin{equation}
\begin{aligned}
B = \frac{4(\nu-\nu_0)^2}{w^2}\log\left[-\frac{1}{\tau}\log\left(\frac{1+e^{-\tau}}{2}\right)\right],
\end{aligned}
\label{eqfgab}
\end{equation}
where $w$ is the full width at half maximum, and $\tau$ is a flattening factor. The unsmooth flat bottom of a flattened Gaussian is what distinguishes it from the  two other types of signal. The parameters used to characterise the flattened Gaussian signal EDGES claimed to have detected have the following values: $A_{21}$ $= 0.52$ K, $\nu_0$ $= 78.3$ MHz, $w$ $= 20.7$ MHz, and $\tau$ $= 6.5$.

\textsc{globalemu} \citep{bevins2021} is a sky-averaged 21-cm signal emulator, and the signal during cosmic dawn and the epoch of reionisation is characterised by the following astrophysical parameters: the star formation efficiency, $f_*$, the minimal virial circular velocity, $V_\mathrm{c}$,  the X-ray efficiency, $f_\mathrm{X}$, the CMB optical depth, $\tau$, the slope of the spectral energy density, $\alpha$, the low energy cut-off of the X-ray spectral energy density, $\nu_\mathrm{min}$, and the mean free path of ionising photons, $R_\mathrm{mfp}$. 
 
Table \ref{tabglob} lists the parameters used to generate the signals adopted in this work. Some signals are emulated with enhanced radio background, where the parameter $\alpha$ is replaced by $f_\mathrm{radio}$, for the deeper absorption troughs that is easier to recover. The models with an enhanced radio background are built to explain the EDGES result \citep{sharmaeb, fiaeb}, the detected signal of which has an amplitude incompatible with the standard 21-cm models. For the purpose of this paper, only signals with absorption trough deeper than 15 mK are used as the true signal.

\begin{table*}
	\centering
	\caption{This is the list of parameters used to generate the \textsc{globalemu} signals tested in this paper. The first four cases are generated using the original version of \textsc{globalemu}, while the last two cases are generated with an enhanced radio background, where the parameter $\alpha$ is replaced by $f_\mathrm{radio}$. The subscript indicates the frequency at which the simulated signal is at its minimum.}
	\label{tabglob}

	\begin{tabular}{lccccccc} 
    		 & $f_*$ & $V_\mathrm{c}$ & $f_\mathrm{X}$ & $\tau$ & $\alpha$ & $\nu_\mathrm{min}$ & $R_\mathrm{mfp}$ \\\hline
    $G_{73}$ & $2.746 \times 10^{-1}$ & $3.795 \times 10^{1}$ & $7.504 \times 10^{1}$ & $3.591 \times 10^{-2}$ & $1.242$ & $2.914$ & $3.262 \times 10^{1}$  \\ 
    $G_{85}$ & $1.715 \times 10^{-1}$ & $5.231 \times 10^{1}$ & $4.756 \times 10^{2}$ & $4.303 \times 10^{-2}$ & $1.188$ & $2.920$ & $2.297 \times 10^{1}$  \\ 
    $G_{99}$ & $3.657 \times 10^{-1}$ & $9.424 \times 10^{1}$ & $2.053 \times 10^{2}$ & $7.732 \times 10^{-2}$ & $1.078$ & $2.814$ & $4.573 \times 10^{1}$  \\ 
    $G_{107}$ & $4.176 \times 10^{-1}$ & $8.783 \times 10^{1}$ & $3.459 \times 10^{1}$ & $7.104 \times 10^{-2}$ & $1.032$ & $2.562$ & $3.842 \times 10^{1}$  \\  
		\hline \\[2mm] 
   & $f_*$ & $V_\mathrm{c}$ & $f_\mathrm{X}$ & $\tau$ & $f_\mathrm{radio}$ &$\nu_\mathrm{min}$ & $R_\mathrm{mfp}$ \\\hline
    $G^*_{73}$ & $4.113 \times 10^{-2}$ & $4.490 \times 10^{1}$ & $1.332 \times 10^{0}$ & $5.753 \times 10^{-2}$ & $2.112 \times 10^{2}$ & $0.000$ & $4.000 \times 10^{1}$ \\ 
    $G^*_{99}$ & $1.588 \times 10^{-1}$ & $1.532 \times 10^{1}$ & $1.096 \times 10^{0}$ & $9.093 \times 10^{-2}$ & $1.129 \times 10^{2}$ & $0.000$ & $4.000 \times 10^{1}$ \\
    
  \hline 
	\end{tabular}
\end{table*}

\subsubsection{Systematics}
In the cases where systematics are present, the simulated antenna temperature data include a sinusoid with the amplitude of \{20 mK, 50 mK, 100 mK\}. The phase of the sinusoids is shifted with respect to the centre frequency of the injected signal or the frequency at which the signal is at its lowest, so the phase difference between the signal and the sinusoid is the same across all cases:
\begin{equation}
\begin{aligned}
T_{\mathrm{sys}} = A_{\mathrm{sys}} \sin{\left(2\pi \frac{(\nu - \nu_0)}{P_{\mathrm{sys}}} + \phi_{\mathrm{sys}} \right)},
\end{aligned}
\label{eqasinsys}
\end{equation}
where $A_{\mathrm{sys}}$ is the amplitude of the sinusoid, $P_{\mathrm{sys}}$ is the period, and $\phi_{\mathrm{sys}}$ is the phase.

\subsubsection{Antenna Temperature Noise}
The antenna temperature noise is generated by the generalized normal distribution noise model where the noise at each frequency channel is randomly assigned based on the probability density distribution. It is a reasonable alternative to the radiometric noise and is easier to generate. The scale parameter of the generalized normal distribution noise model can be expressed as 
\begin{equation}
\begin{aligned}
\alpha_\mathrm{n} = \sigma_\mathrm{n} \sqrt{\frac{\Gamma\left(\frac{1}{\beta_\mathrm{n}}\right)}{\Gamma\left(\frac{3}{\beta_\mathrm{n}}\right)}},
\end{aligned}
\label{eqgnorm}
\end{equation}
where $\beta_\mathrm{n}$ is the shape parameter, $\sigma_\mathrm{n}$
is the standard deviation, and $\Gamma$ denotes the gamma function. In this work, the shape parameter is set to  $\beta_\mathrm{n} = 2$, and standard deviation is $\sigma = 25$ mK unless otherwise specified.

\subsection{Foreground Modelling and Fits}
In the REACH data analysis pipeline, a foreground model, $\mathcal{M}_\mathrm{f}$, and a signal model, $\mathcal{M}_\mathrm{21}$, are jointly fitted; with the noise model, $\mathcal{M}_{\mathrm{noise}}$, together they work as a single model $\mathcal{M}$ to be analysed by the Bayesian algorithm (Eq. \ref{eqmn}). The foreground modelling follows the same process as how the foreground component of the simulated data is generated (section \ref{sec321}). Moreover, in order to take chromatic distortions into account when modelling the foregrounds, the sky is first divided into $N$ regions of similar spectral indices, and in each region, a distinct uniform spectral index parameter is assigned for the purpose of scaling the base map \citep{anstey}. The antenna temperature can then be modelled by convolving it with a beam pattern. Chromatic distortions due to beam chromaticity and non-uniform spectral index can be characterised by fitting this physically motivated foreground function. The foreground model, $\mathcal{M}_\mathrm{fg}$, is then fit to the data jointly with the signal model, $\mathcal{M}_\mathrm{21}$, via Bayesian inference with a likelihood function of 
\begin{equation}
\begin{aligned}
\log \mathcal{L} = \sum_i \log \left( 2 \pi \sigma_\mathrm{n}^2 \right) - \frac{1}{2} \left( \frac{T_\mathrm{data}(\nu_i)-(T^*_\mathrm{fg}(\nu_i)+T^*_\mathrm{21}(\nu_i))}{\sigma_\mathrm{n}}\right)^2,
\end{aligned}
\label{eqlike}
\end{equation}
where $i$ is the index of each frequency bin, under the assumption of a simple model of uniform uncorrelated Gaussian noise $\sigma_\mathrm{n}$ across the frequency band. $T^*_\mathrm{fg}$ and $T^*_\mathrm{21}$ are the antenna temperature data given by the models $\mathcal{M}_\mathrm{fg}$ and $\mathcal{M}_\mathrm{21}$ respectively. The prior given to the spectral index parameters $\beta_i$ is [$2.45844,3.14556$] (uniform prior), which is the full range of spectral indices in the map, and  the Gaussian noise parameter [$10^{-4}, 10^1$] K (logarithmically uniform prior). The \textsc{PolyChord} settings used in this paper are listed in Table \ref{tabpoly}.

\begin{table}
	\centering
	\caption{\textsc{PolyChord} settings applied in all the tests performed in this paper. nDims $=n_\mathrm{foreground} + n_\mathrm{signal} + 1$ is the dimensionality of the model, where $n_\mathrm{foreground}$ is the number of regions, set to 9 in this paper, and $n_\mathrm{signal}$ is the number of parameters used in the signal model. The final parameter is the uncorrelated Gaussian noise, $\sigma_\mathrm{n}$.}
	\label{tabpoly}
	\begin{tabular}{ll} 
		Parameter             & Setting    \\\hline
        $n_\mathrm{live}$                 & nDims * 25  \\  
		$n_\mathrm{repeats}$           & nDims * 5    \\ 
		$n_\mathrm{prior}$                & nDims * 25  \\  
        $n_\mathrm{fail}$               & nDims * 25  \\    
        do clustering       & True        \\
        precision criterion & 0.001        \\
		\hline 
	\end{tabular}
\end{table}
 
\section{Results}
\label{sec4}
\subsection{Model Comparison}
In theory, the FlexKnot signal model should have the versatility to fit injected signals of different types of shape, and for the cubic spline twice integrated from linear spline method, shapes that are smooth without rigid turns are favoured. Signals of different types of shape, namely, the Gaussian signal, the flattened Gaussian signal, and the \textsc{globalemu} signal, are injected in the data to test the model's capabilities and make comparison with the commonly used Gaussian signal model. Apart from $\Delta \log(\mathcal{Z})$ (Eq. \ref{eqmodelprefer}), signal root-mean-square error (RMSE), the RMSE between the injected signal, $T_{\mathrm{21}}$, and the reconstructed signal, $T^*_{\mathrm{21}}$, is also calculate to compare between different models:
\begin{equation}
\begin{aligned}
\mathrm{RMSE} = \sqrt{\sum_{i=1}^{n}\frac{(T_{\mathrm{21}}(\nu_i) - T^*_{\mathrm{21}}(\nu_i))^2}{n}}.
\end{aligned}
\label{eqrmse}
\end{equation}
A lower signal RMSE means a better fit.

\subsubsection{Gaussian}
\begin{figure*}
    \centering
    \minipage{0.96\textwidth} \includegraphics[trim={0cm 0cm 0 0cm},clip,width=\linewidth]{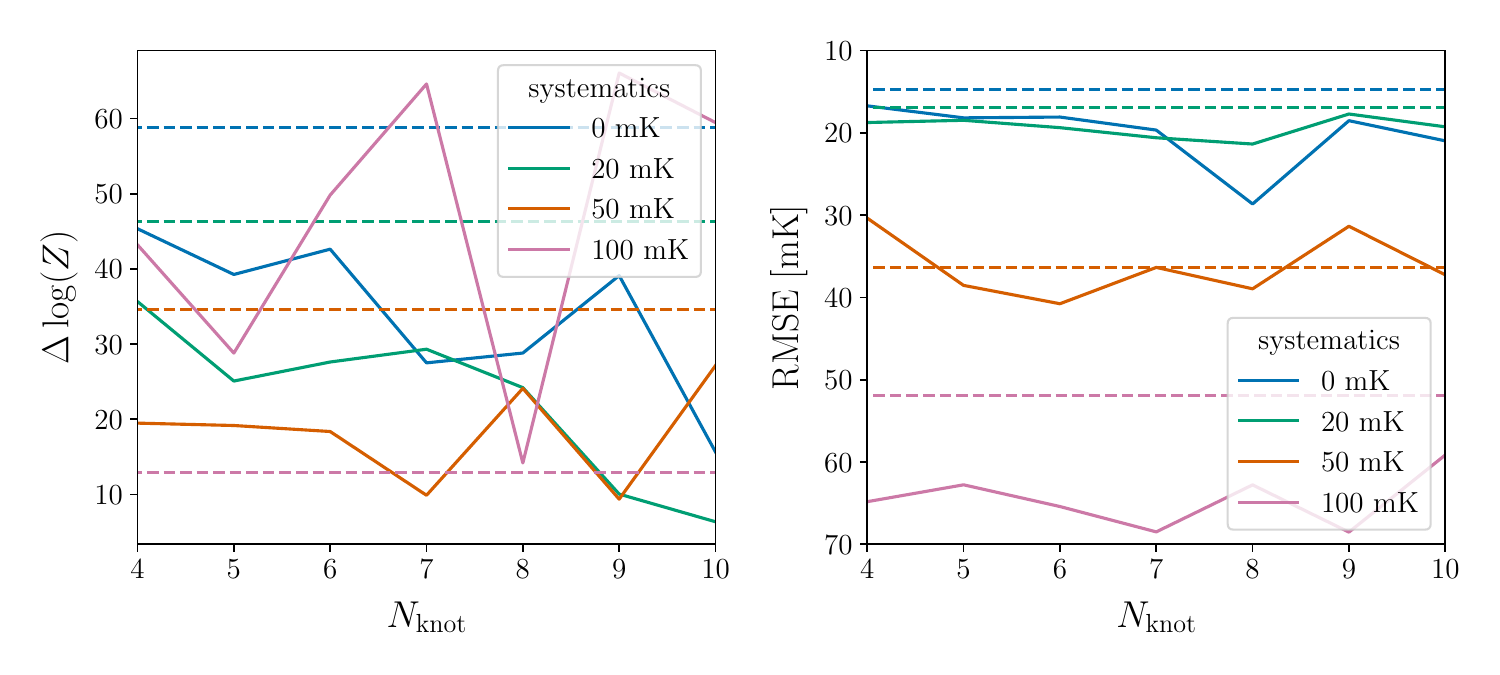}
    \endminipage\hfill
    \caption{Gaussian and FlexKnot signal model comparison in $\Delta \log(\mathcal{Z})$ and the signal RMSE between the recovered signal and the simulated signal. The true signal is a Gaussian centred at 85 MHz with an amplitude of 155 mK and a standard deviation of 10 MHz. The results yielded by the FlexKnot model are shown in solid lines, and the Gaussian cases are shown in dashed lines as reference. The left panel shows high $\Delta \log(\mathcal{Z})$ in all cases, indicating that the model with a signal is strongly preferred. Moreover, the overall decreasing $\Delta \log(\mathcal{Z})$ with $N_\mathrm{knot}$ shows that a higher number of knots does not necessarily improve the confidence in signal recovery. The y-axis is inverted in the right panel. In the right panel, it can be seen that the signal RMSE's are lower when the Gaussian signal model is adopted, which means the FlexKnot signal model does not yield better signal recoveries than the Gaussian signal model. The signal RMSE yielded by the FlexKnot model is nevertheless only slightly higher the Gaussian fit in the cases with low level systematics. The signal RMSE panel also shows that the presence of systematic decreases the quality of fit progressively with the level of its amplitude in all cases, with the case of 50 mK and 100 mK having significantly higher values, indicating the unreliability of the signal recovery.}
    \label{figgauss}
\end{figure*}

In the tested cases where the injected signal is a Gaussian (the tested parameter space is listed in section \ref{secsignal}), the FlexKnot model usually does not recover the signal better than the Gaussian model itself. In most of the cases (Table \ref{tabcom} is the list of all tested cases), the FlexKnot model nevertheless recovers the signals with significant confidence in terms of $\Delta \log(\mathcal{Z})$ and with the signal RMSE being only slightly higher than that of the Gaussian fit. Tested cases with an injected Gaussian signal yield very similar results, and Fig. \ref{figgauss} shows one of the cases where the true signal is a Gaussian centred at 85 MHz with an amplitude of 155 mK and a standard deviation of 10 MHz. The dashed lines in the figure are the corresponding reference values yielded by the Gaussian signal model, and the solid lines are that of the FlexKnot model of different $N_\mathrm{knot}$. The signal RMSE's are lower than when using the Gaussian signal model, indicating better signal fits. The signal RMSE also shows that the presence of systematic decreases the quality of fit progressively with the level of its amplitude in all cases, with the case of 50 mK and 100 mK having significantly higher values, indicating the unreliability of the signal recovery. The impact of systematics is discussed more extensively in section \ref{secsys}. 

\begin{table}
	\centering
	\caption{Signal RMSE and $\Delta \log(\mathcal{Z})$ comparison between the FlexKnot model $\mathcal{M}_\mathrm{FK}$ and the Gaussian model $\mathcal{M}_\mathrm{G}$. The injected signal is in the shape of a Gaussian. $\mathcal{M}_\mathrm{G}$ yields a lower signal RMSE in most of the cases and a higher $\Delta \log(\mathcal{Z})$ in all of the listed cases.}
	\label{tabcom}

	\begin{tabular}{ccc|cc|cc} 
    \multicolumn{3}{c}{{signal parameter}} & \multicolumn{2}{c}{\textbf{RMSE}} & \multicolumn{2}{c}{\bm{$\Delta \log(\mathcal{Z})$}} \\ \hline
    $\nu_0$ & $\sigma$ & $A_{21}$ & $\mathcal{M}_\mathrm{FK}$ & $\mathcal{M}_\mathrm{G}$ & $\mathcal{M}_\mathrm{FK}$ & $\mathcal{M}_\mathrm{G}$ \\ \hline
    85 & 10 & 155 & 16.7 & 14.8 &  45.3 & 58.7 \\
    85 & 15 & 155 & 15.8 & 15.9 &  41.0 & 61.1 \\
    85 & 20 & 155 & 22.6 & 19.8 &  39.7 & 52.0 \\
    105 & 10 & 155 & 25.1 & 23.9 &  78.6 & 94.0 \\ 
    105 & 15 & 155 & 18.9 & 32.2 &  78.6 & 94.9 \\
    105 & 20 & 155 & 10.5 & 14.2 &  68.7 & 84.3 \\
    125 & 10 & 155 & 14.6 & 2.3 &  85.7 & 91.7 \\
    125 & 15 & 155 & 11.8 & 7.9 &  103.5 & 114.5 \\
    125 & 20 & 155 & 8.0 & 3.0 &  107.2 & 119.7 \\
    85 & 10 & 255 & 21.3 & 11.7 & 90.0 & 96.0 \\
    85 & 15 & 255 & 20.8 & 25.4 &  87.1 & 95.0 \\
    85 & 20 & 255 & 51.9 & 34.2 &  65.9 & 85.8 \\
    105 & 10 & 255 & 30.9 & 11.3 &  127.0 & 145.6 \\
    105 & 15 & 255 & 24.6 & 5.7 &  133.2 & 142.9 \\
    105 & 20 & 255 & 10.4 & 7.3 &  128.3 & 137.2 \\
    125 & 10 & 255 & 14.6 & 3.6 &  143.3 & 154.4 \\
    125 & 15 & 255 & 19.6 & 5.3 &  170.6 & 177.6 \\
    125 & 20 & 255 &  18.7 & 5.4 &  176.4 & 184.6 \\
    85 & 10 & 455 & 24.6 & 14.3 & 158.4 & 164.7 \\
    85 & 15 & 455 & 27.4 & 18.6 & 142.0 & 162.2 \\
    85 & 20 & 455 & 44.4 & 26.0 &  129.1 & 146.8 \\
    105 & 10 & 455 & 49.0 & 21.1 & 220.5 & 239.4 \\
    105 & 15 & 455 & 19.2 & 28.4 & 225.4 & 244.2 \\
    105 & 20 & 455 & 17.6 & 12.7 & 217.2 & 232.7 \\
    125 & 10 & 455 & 17.1 & 2.5 & 245.4 & 261.5 \\
    125 & 15 & 455 & 29.4 & 4.4 & 306.8 & 313.6 \\
    125 & 20 & 455 & 21.3 & 3.1 & 330.3 & 346.0  \\
  \hline
	\end{tabular}
\end{table}

\subsubsection{\textsc{globalemu}}
Within the tested parameter space (listed in Table \ref{tabglob}), the FlexKnot model performs better than the Gaussian model in the cases of injected \textsc{globalemu} signal (Table \ref{tabcomgem}); it is able to recover various absorption histories of the \textsc{globalemu} signal yielded by different set of astrophysical parameters. The parameters used to generate the \textsc{globalemu} signals tested in the work are shown in Table \ref{tabglob}.

\begin{figure*}
    \centering
    \minipage{0.96\textwidth} \includegraphics[trim={0cm 0cm 0 0cm},clip,width=\linewidth]{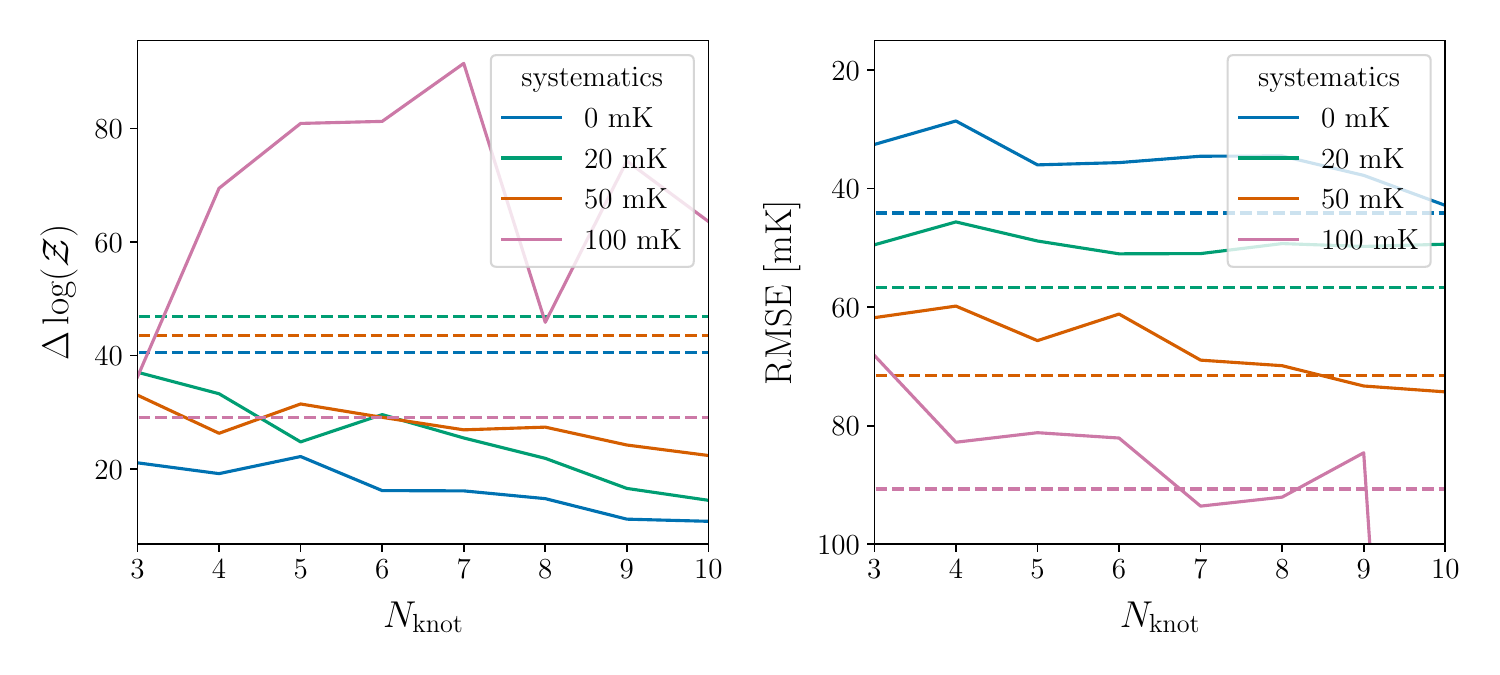}
    \endminipage\hfill
    \caption{Gaussian and FlexKnot signal model comparison in $\Delta \log(\mathcal{Z})$ and the signal RMSE between the recovered signal and the simulated signal. The true signal is a \textsc{globalemu} signal whose minimum point is at 73 MHz with an amplitude of about 150 mK. The parameters used to generate the signal are listed in the first entry $G_{73}$ of Table \ref{tabglob}. The results yielded by the FlexKnot model is shown are solid lines, and the Gaussian cases are shown in dashed lines as reference. The left panel shows that $\Delta \log(\mathcal{Z})$ is sufficiently high in all cases, indicating that the model with a signal is strongly preferred. The decreasing $\Delta \log(\mathcal{Z})$ shows that a higher number of knots does not improve the confidence in signal recovery. The y-axis is inverted in the right panel. The right panel shows that in these cases, by the lower signal RMSE's, the FlexKnot signal model yields better signal recoveries than the Gaussian signal model, and 4 to 6 knots is optimal and sufficient to recover the signal. Alike to the Fig. \ref{figgauss}, it shows that the presence of systematic decreases the quality of fit progressively with the level of its amplitude in all cases, with the case of 50 mK and 100 mK having high values, indicating the unreliability of the signal recovery.}
    \label{figgvsf}
\end{figure*}

\begin{figure}
    \centering
    \minipage{0.49\textwidth} \includegraphics[trim={0cm 0cm 0 0cm},clip,width=\linewidth]{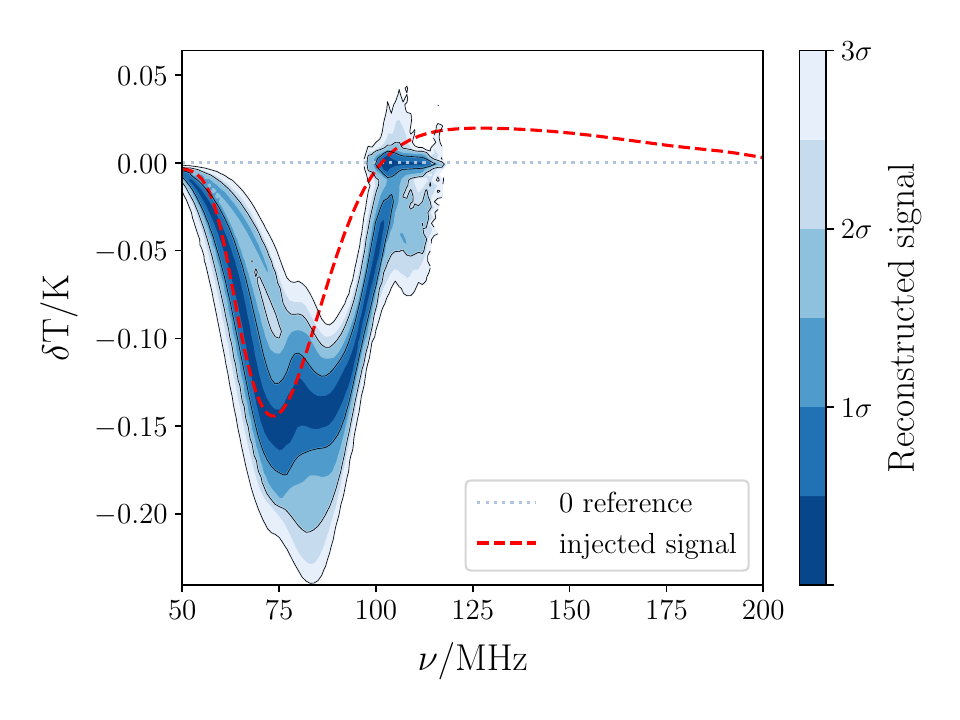}
    \endminipage\hfill
    \caption{An example of the reconstructed signal (predictive posterior of the final function given the distribution of the parameter posteriors) using a 4-knot FlexKnot signal model where the injected signal is a \textsc{globalemu} signal ($G_{73}$ in table \ref{tabglob}) with significant emission at higher frequencies. The emission at higher frequencies is not captured by the signal model because the value is set to be zero after the highest frequency knot.}
    \label{figglobg4}
\end{figure}

One of the results is shown in Fig. \ref{figgvsf}. The high $\Delta \log(\mathcal{Z})$ indicates the significant confidence in the results of all the explored cases. The signal RMSE shows a significant difference between the FlexKnot model and the Gaussian model, the former outperforming the latter. The signal RMSE may seem high for an injected signal at the absorption level of $A \sim 145$ mK. It is due to the continuous significant emission at the higher frequencies in this particular case, which works against the FlexKnot model whose values after the highest frequency knot are set to zero. This is shown in Fig. \ref{figglobg4}. The discrepancies at the high frequency range contribute significantly to the signal RMSE. If the higher frequency (emission) range is excluded, the signal RMSE becomes much lower.

\begin{table}
	\centering
	\caption{Signal RMSE and $\Delta \log(\mathcal{Z})$ comparison between the FlexKnot model $\mathcal{M}_\mathrm{FK}$ and the Gaussian model $\mathcal{M}_\mathrm{G}$. The injected signal is generated by \textsc{globalemu}. $\mathcal{M}_\mathrm{FK}$ yields a lower signal RMSE in all of the listed cases. The subscript indicates the frequency at which the simulated signal is at its minimum.}
	\label{tabcomgem}

	\begin{tabular}{ccc|cc} 
    & \multicolumn{2}{c}{\textbf{RMSE}} & \multicolumn{2}{c}{\bm{$\Delta \log(\mathcal{Z})$}} \\ \cline{2-5}
     & $\mathcal{M}_\mathrm{FK}$ & $\mathcal{M}_\mathrm{G}$ & $\mathcal{M}_\mathrm{FK}$ & $\mathcal{M}_\mathrm{G}$ \\ \hline
     $G_{73}$ & 28.6 & 44.1 & 22.2 & 40.6 \\
     $G_{85}$ & 42.9 & 46.9 & 24.9 & 43.5 \\
     $G_{99}$ & 17.4 & 25.7 & 53.9 & 69.6 \\
     $G_{107}$ & 21.4 & 35.5 & 71.5 & 93.1 \\
     
  \hline
	\end{tabular}
\end{table}

\subsubsection{Flattened Gaussian}
\begin{figure}
    \centering
    \minipage{0.5\textwidth} \includegraphics[trim={0.99cm 0cm 0.99cm 0cm},clip,width=\linewidth]{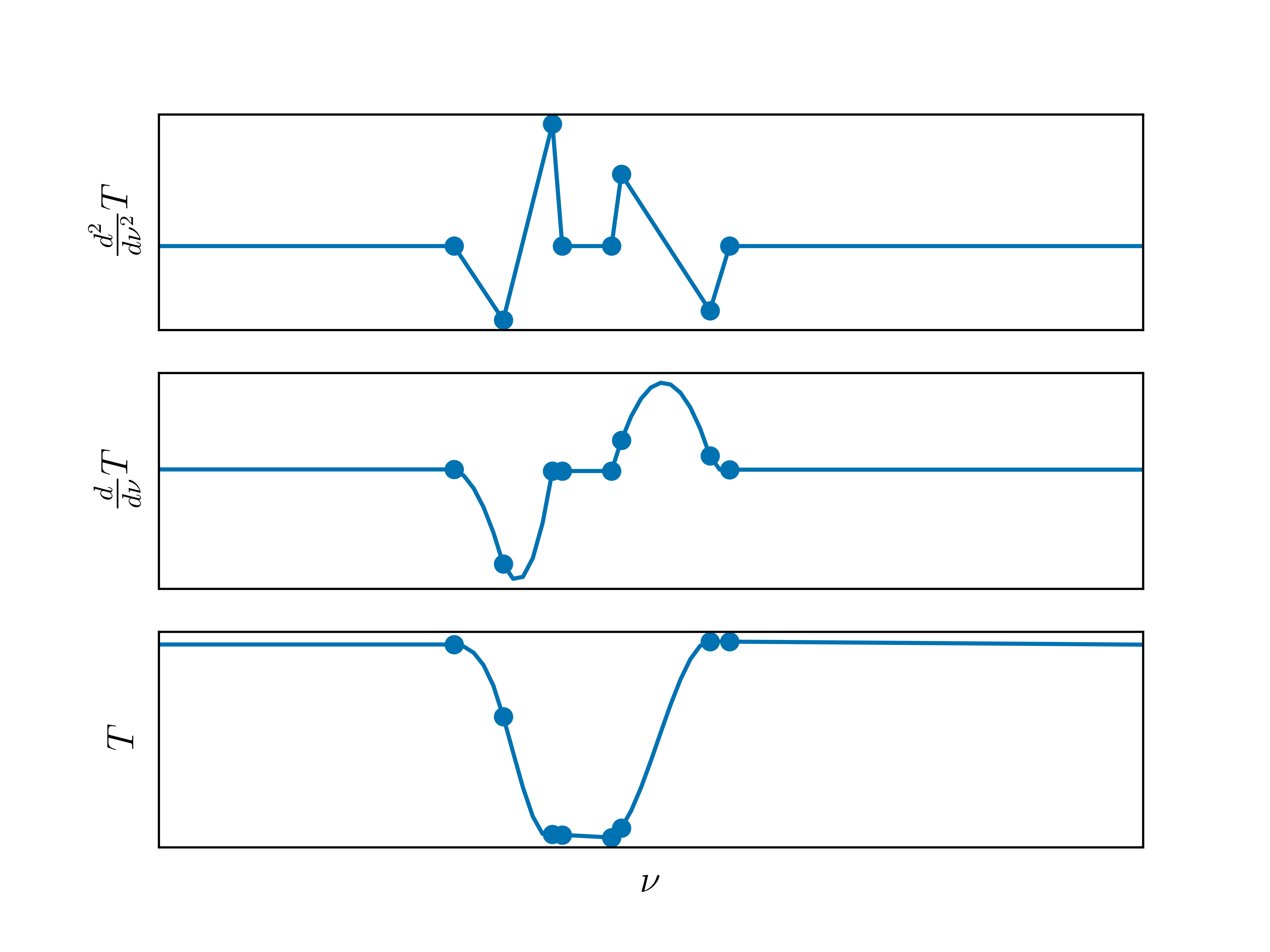}
    \endminipage\hfill
    \minipage{0.49\textwidth} \includegraphics[trim={0cm 0cm 0cm 0cm},clip,width=\linewidth]{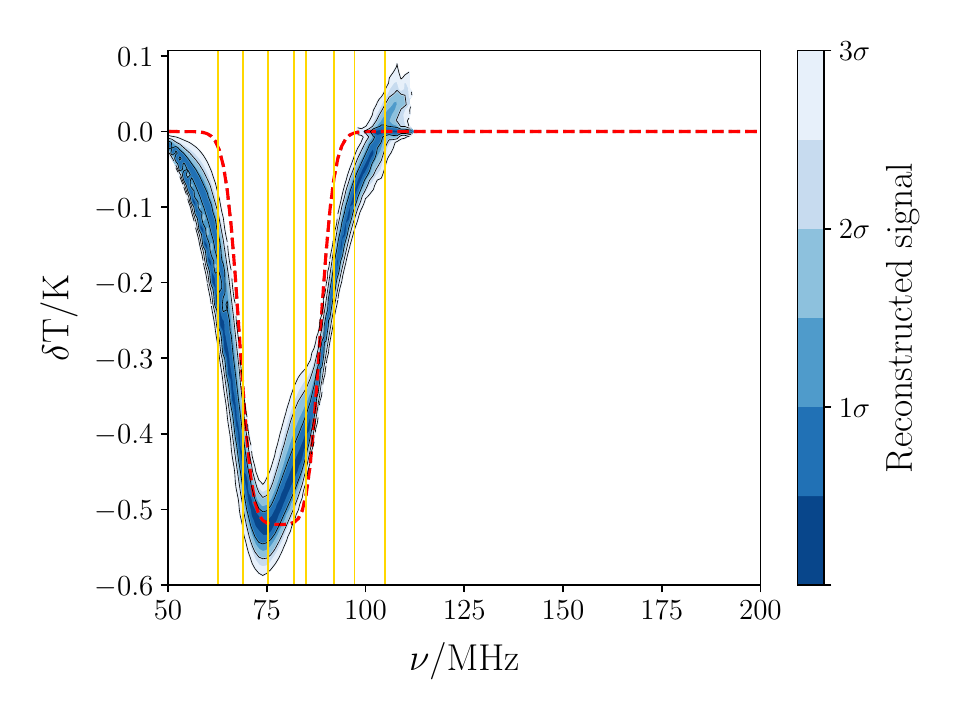}
    \endminipage\hfill
    \caption{The upper panels, like Fig. \ref{figsd}, show how the FlexKnot function is implemented in our signal model in three steps, except that the signal has the shape of a flattened Gaussian: the second derivative of the global 21-cm signal is parameterised by linearly interpolated splines, shown in the first panel, and the signal is recovered by integrating it twice, shown respectively in the second and the third panel. It shows the minimum number of knots to fully describe a flattened Gaussian. The contour plot (predictive posterior of the final function given the distribution of the parameter posteriors) in the bottom shows the reconstructed signal in the case where the true signal is a flattened Gaussian. The yellow lines indicate the mean value of the location of each knot when $N_{\mathrm{knots}} = 7$. The knots do lie in the position similar to what the upper panels suggest, but there are only three knots in the bottom, one knot short of the requirement to form a flat bottom. Similar results are yielded even when using a higher number of knots.}
    \label{figsdfl}
\end{figure}

Implementing flattened Gaussian shaped global signals is motivated by \citet{edges}. The shape of a flattened Gaussian is different from that of a Gaussian or a \textsc{globalemu} in its flat bottom, making the overall shape resemble piecewise linear spline more than piecewise cubic spline. In the cases of injected flattened Gaussian signals, the centre frequency is well recovered for both models. The sharp corners at the bottom of the flattened Gaussian already rule out the possibility for the Gaussian model to recover it accurately. On the other hand, as shown in Fig. \ref{figsdfl}, the FlexKnot model is able to accurately capture the width; it recovers the shape well up until the bottom of the signal, where the Gaussian is flattened. The sharp corners of the signal, as well as the flat bottom, however, are not. In the presence of the chromatic foregrounds, it often fails to capture the feature using the necessary four knots, but instead uses fewer knots to describe a smoother shape. When there are more than sufficient knots, the model would begin to fit structures such as the foreground chromaticity. The reconstructed signal plot at the bottom of Fig. \ref{figsdfl}, in which the yellow vertical lines indicate the mean value of each knot location yielded by the fit when $N_{\mathrm{knots}} = 7$, shows that the location of most knots are not too far-off, but it is one knot short of the four knots necessary to form a flat bottom. It is partly due  to the high noise level $\sigma=25$ mK; the recovery of the flattened bottom could sometimes be improved by lowering the noise level in $T_\mathrm{data}$. An alternative is to simply adopt linear spline interpolation instead of cubic spline in the FlexKnot signal model. 

\subsection{Signal Recovery with Systematics}
\label{secsys}
\begin{figure}
    \centering
    \minipage{0.5\textwidth} \includegraphics[trim={0cm 2.1cm 0cm 0.4cm},clip,width=\linewidth]{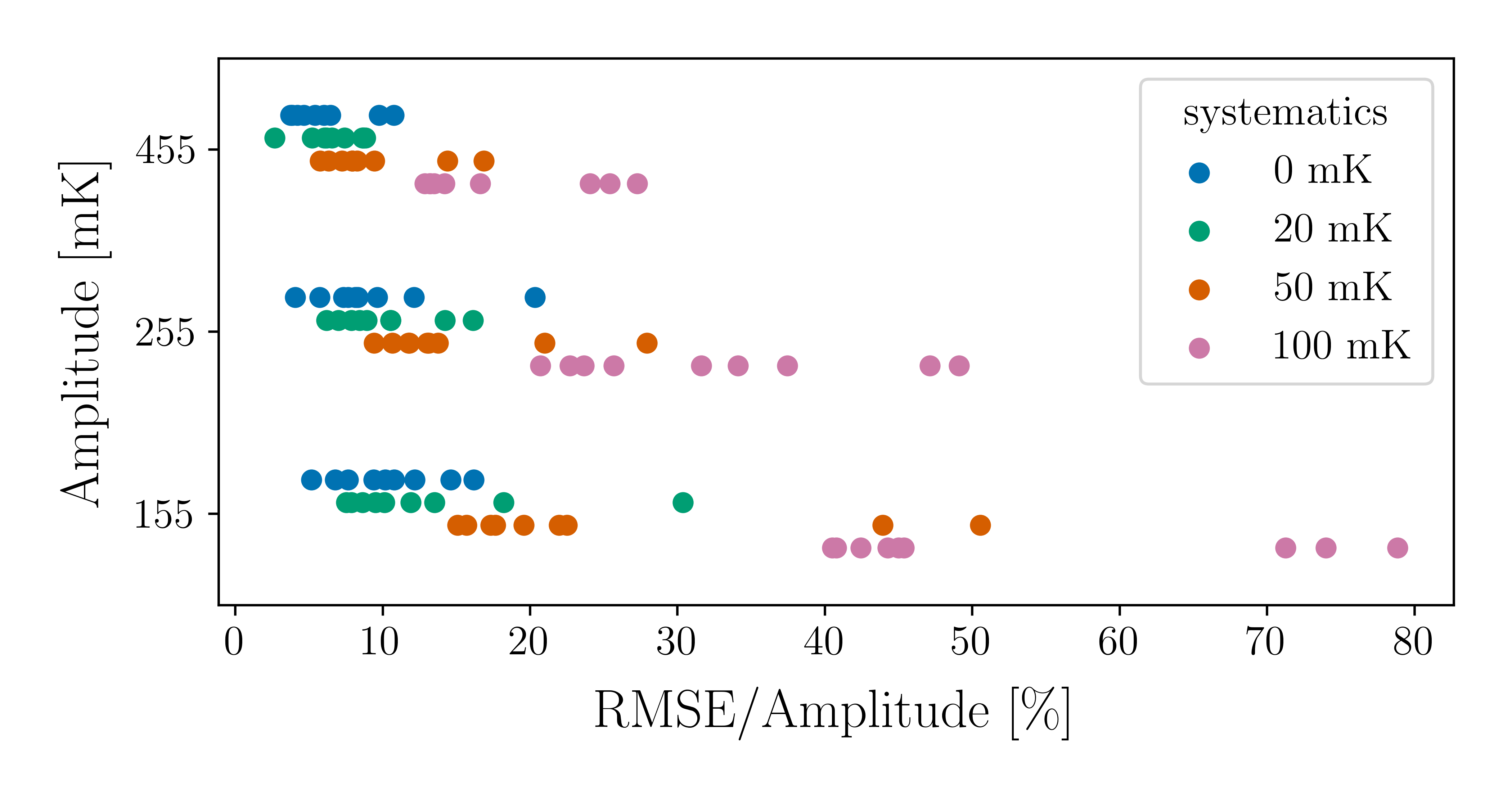}
    \endminipage\hfill
    \minipage{0.5\textwidth} \includegraphics[trim={0cm 2.1cm 0cm 0.4cm},clip,width=\linewidth]{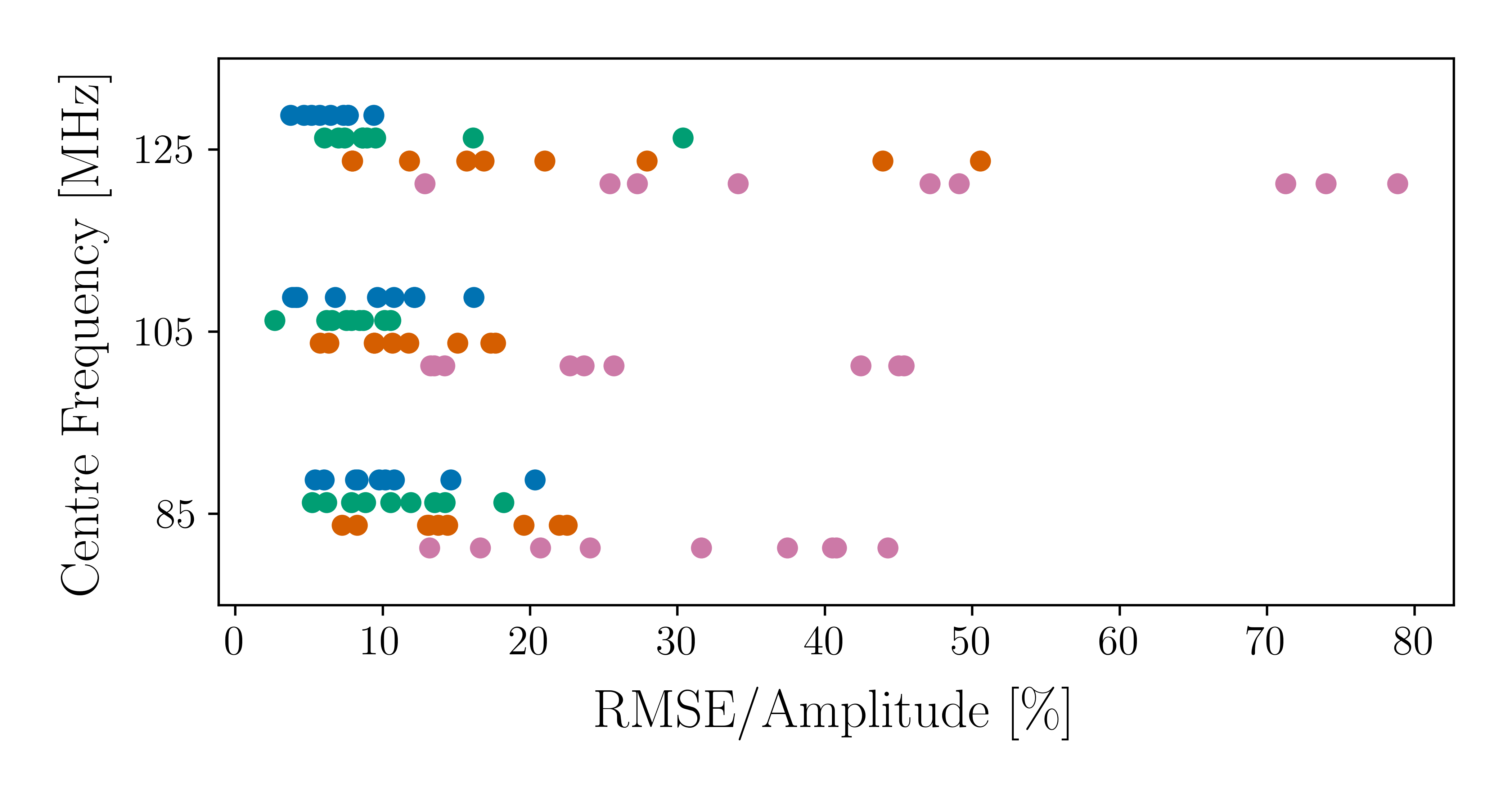}
    \endminipage\hfill
    \minipage{0.5\textwidth} \includegraphics[trim={0cm 0cm 0cm 0.4cm},clip,width=\linewidth]{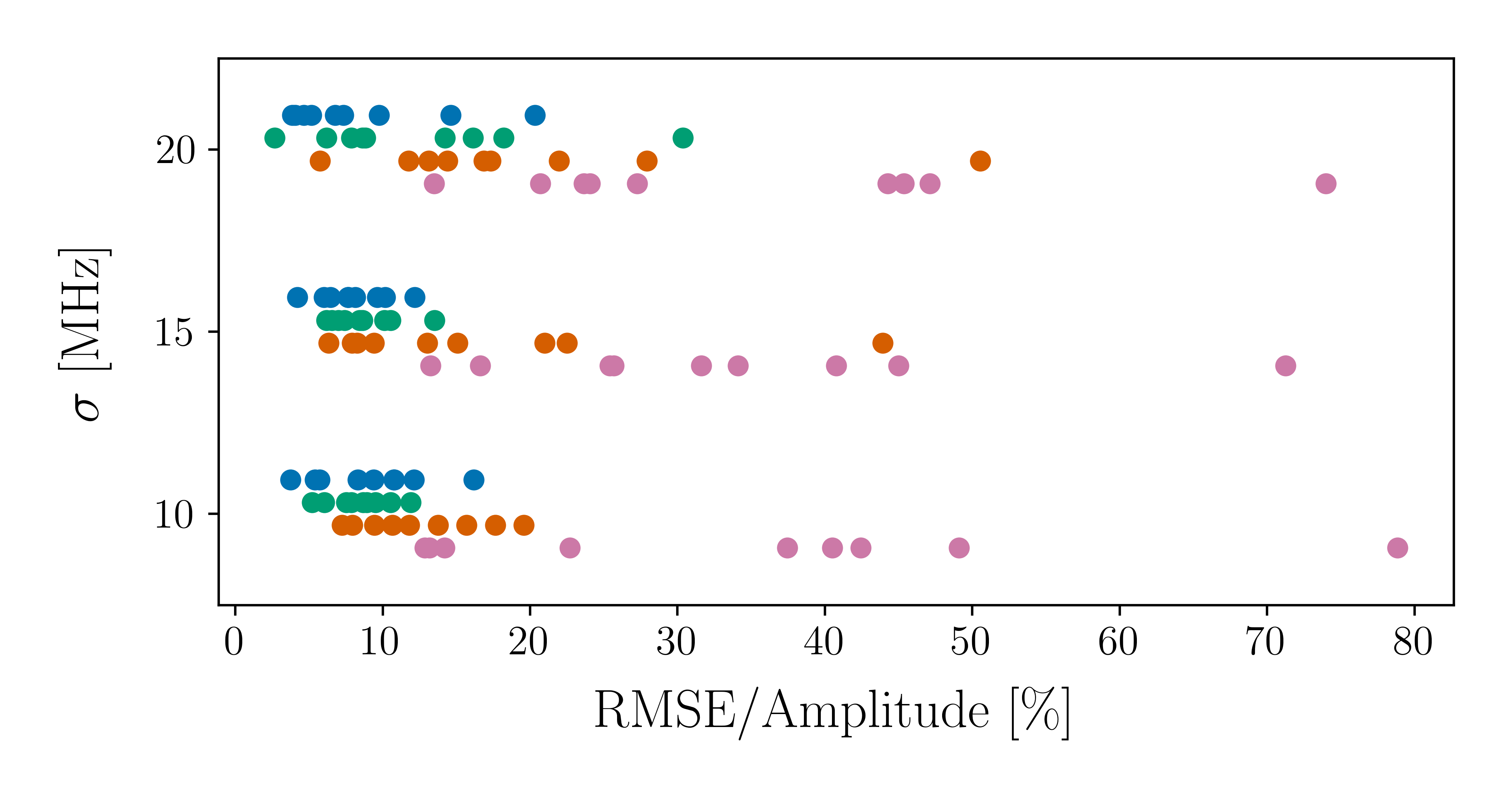}
    \endminipage\hfill
    \caption{Signal RMSE divided by the amplitude of the injected Gaussian signal of various different cases with or without systematics, as indicated by the legend.  Different colours indicate the level of the added systematics. All three panels show the same set of data. Together it shows a full 3D parameter sweep that is marginalised onto different parameters individually. From top to bottom, the cases are grouped respectively by the amplitude, the centre frequency, and the standard deviation of the injected Gaussian signal.  The first panel shows that signal recovery is generally better when the true signal has a deeper absorption level. The second panel shows that without systematics, signals at higher centre frequency yield slightly better recovery, but the advantage no longer exists in the presence of the systematics. The third panel does not show any apparent difference signals with different standard deviation. Overall, within the explored parameter space, the quality of the signal recovery is no longer reliable when the sinusoid systematics reach 50 mK.}
    \label{figrmseamp}
\end{figure}

In order to explore the capability of the FlexKnot signal model in the presence of systematics, simulated antenna temperature data with different levels of sinusoids (at an amplitude of \{0 mK, 20 mK, 50 mK, 100 mK\}) and injected signals have been tested. Each case is tested using a range of $N_{\mathrm{knots}}$, of which the lowest signal RMSE is recorded. To compare the fit of the signals at different absorption levels, the signal RMSE is divided by the amplitude of the true signal, yielding a dimensionless value. The results are shown in Fig. \ref{figrmseamp}, where the cases are grouped by the amplitude, the centre frequency, and the standard deviation of the injected signal.

In all cases, as expected, greater systematics in $T_\mathrm{data}$ would yield higher signal RMSE, or poorer fits. The first panel is grouped by the amplitude of the true signal, showing that the stronger the signal absorption level, the less susceptible it is to sinusoidal systematics, and thus the better the signal recovery is.

The second panel, grouped by the centre frequency of the true signal, shows that when there is no systematic present, signal recovery is slightly better when the true signal is centred at a higher frequency. The foreground brightness decreases exponentially with frequency, and thus the signal presence becomes more distinctive at higher frequencies, the likely reason for the better results. It no longer holds, however, in the presence of sinusoidal systematics, which, depending on its phase, can distort the shape of the signal and cause a shift in the location of the signal trough.

There is no apparent difference between the groups by the standard deviation. It probably is due to the period of the added sinusoidal systematics, which is larger than the entirety of the explored parameter space covering only \{10, 15, 20\} MHz.  

Overall, the results show that antenna temperature data with sinusoidal systematics of an amplitude higher than 50 mK could not be recovered with sufficient accuracy using the FlexKnot signal model.

\subsection{Optimal Number of Knots}
In this section, we investigate the optimal number of knots that could represent the global 21-cm signal. This is done by running the algorithm using a range of number of knots, $N_{\mathrm{knot}}= 4-10$, and examine the fitted signals by log evidence and signal RMSE.
\subsubsection{Data without Systematics}
\begin{figure*}
    \centering
    \minipage{\textwidth} \includegraphics[trim={0cm 0.4cm 0cm 0.4cm},clip,width=\linewidth]{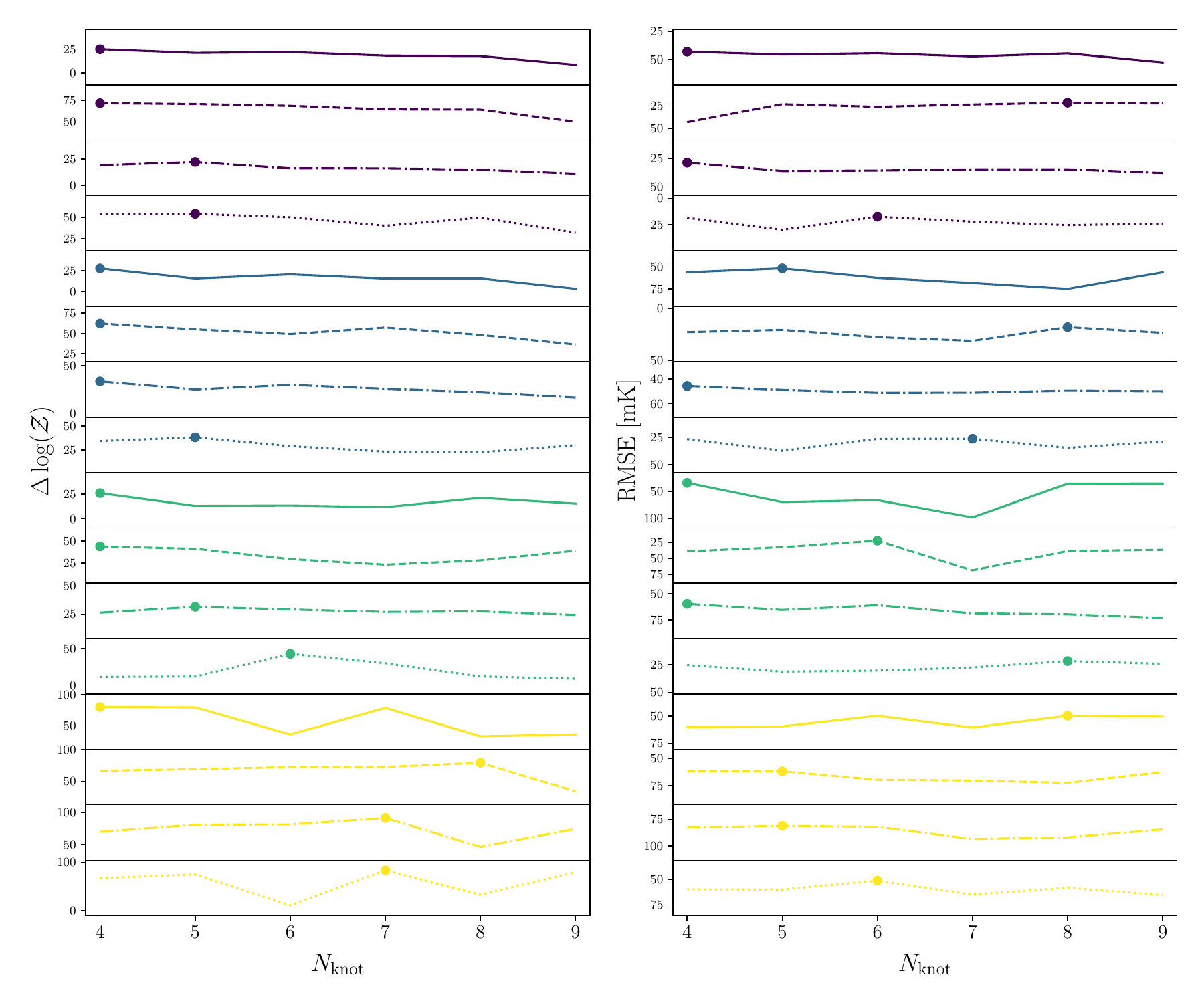}\\[2.5mm]
    \centering
    \includegraphics[width=0.9\linewidth]{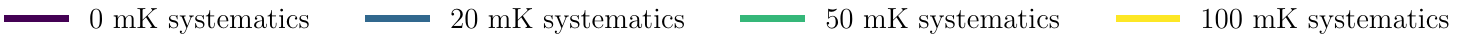}\\[1.5mm]
    \centering
    \includegraphics[width=0.45\linewidth]{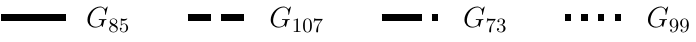}\\[1.5mm]
    \endminipage\hfill
    \caption{Dependence of $\Delta \log(\mathcal{Z})$ (left) and the signal RMSE (right) on the number of knots of multiple different injected signals generated by \textsc{globalemu} (parameters used to generate the signals are listed in Table \ref{tabglob}). The colours represent different levels of sinusoidal systematic in the data and the line styles represent different injected \textsc{globalemu} signals, as indicated by the legend in the bottom of the figure. The y-axis showing the signal RMSE on the right hand side has been inverted, meaning what appears to be higher in the plot has a lower RMSE value, and the lowest point appears as a peak in the figure. The highest $\Delta \log(\mathcal{Z})$ and the lowest signal RMSE of each case are marked in dots. Overall, the evidence decreases with increasing number of knots after its peak at $N_{\mathrm{knot}}\sim 4-6$, meaning a higher $N_{\mathrm{knot}}$ does not significantly improve the signal fit. The signal RMSE is at its lowest at $N_{\mathrm{knot}}\sim 4-6$ in the majority of cases.}
    \label{figrmseall}
\end{figure*}

The optimal number of knots in general is the same with or without the presence of foregrounds and noises for both Gaussian and \textsc{globalemu} signals. In the cases without sinusoidal systematics in $T_\mathrm{data}$, $\Delta \log(\mathcal{Z})$ mostly peaks at $N_{\mathrm{knot}}\sim 4-6$ and in general decreases with the increasing $N_{\mathrm{knot}}$, meaning a higher $N_{\mathrm{knot}}$ does not necessarily contribute significant improvement to the signal recovery. This is true for both injected Gaussian and \textsc{globalemu} signals, and it means that $N_{\mathrm{knot}}\sim 4-6$ is sufficient to recover signals in the form of a Gaussian or \textsc{globalemu}. Fig. \ref{figrmseall} shows the signal RMSE to $N_{\mathrm{knot}}$ of multiple different injected \textsc{globalemu} signals. The reconstructed signal plots as well as signal RMSE also show that more knots does not necessarily improve the signal recovery, and instead introduces unwanted structures to the signal fit. Interestingly, despite the different implementations and interpolation methods, our conclusion with regards to the optimal number of knots for data without systematics is similar to \citet{heimer23}, in which it is shown that the evidence peaks at $N_{\mathrm{knot}}=6$ using EDGES low-band data.

On the other hand, the optimal number of knots is different with or without foregrounds and noises for a flattened Gaussian signal. When the data consists of a flattened Gaussian signal only, the total number of knots required to resolve the flattened bottom would be $N_{\mathrm{knot}}\sim 7-8$. When the chromatic foregrounds and noises are present, however, it is often the case that the flat bottom becomes unrecoverable, regressing the signal fit to a smoother bottom, the degree of which depends on the noise level. When the flat bottom is masked by the noise, the log evidence peaks at $N_{\mathrm{knot}}\sim 5-6$, similar to a Gaussian or a \textsc{globalemu} signal.

\subsubsection{Data with Systematics}
When the systematics are at the level where the signal can still be recovered with significant confidence and low signal RMSE, mostly when $A_\mathrm{sys} < 50$ mK, the optimal $N_\mathrm{knots}$ is about the same as the cases without systematics in $T_\mathrm{data}$. Additional knots beyond that tend to start fitting the systematics mixed with the foreground residuals instead of improving the overall signal recovery. $\Delta \log(\mathcal{Z})$ does not improve with the increasing $N_\mathrm{knots}$ either.

In the cases where there are large sinusoidal systematics ($A_\mathrm{sys} > 50$ mK) in the data  that the signal RMSE is high, one sometimes observes a significant increase in $\Delta \log(\mathcal{Z})$ with increasing $N_{\mathrm{knot}}$. This is still a result of the FlexKnot signal model starting to fit the sinusoidal systematics alongside the foreground residuals instead of the true signal; it is confidently fitting the other structures such as part of the foregrounds or the systematics other than the true signal in the given data.

\section{Conclusions}
\label{seccon}
In this work we have explored the capability of a physics-agnostic FlexKnot signal model implemented in the data analysis pipeline of the REACH experiments to recover the cosmological global 21-cm signal from simulated data. We examined the performance of the signal model with and without systematics, as well as for several different implementations of the injected global signal.

Our FlexKnot signal model is characterised by a cubic function integrated twice from a function of freely moving knots interpolated by piecewise linear splines, a different implementation of the FlexKnot signal model  from previous works (e.g. \citet{heimer23}). It is conditioned by the dark ages primer outside the REACH observing frequency range ($50-200$ MHz, which covers the period between cosmic dawn and reionisation) at the lower frequency end. The second derivative prior of the first knot is constrained to be negative to prevent the potential surge of the signal, which would suggest an emission at the lower frequency end corresponding to higher redshifts that is against established theoretical predictions. An additional highest frequency knot is also included to set the ensuing values at higher frequencies to zero.

We implement the FlexKnot signal model in the REACH data analysis pipeline, where the foreground and the signal are jointly fitted, to test how well the signal model manages to recover the injected signal. For comparison, the same tests are performed using the Gaussian signal model. Three types of global 21-cm signal have been used as the true signal: the Gaussian signal, the flattened Gaussian signal, and physically motivated signals emulated by \textsc{globalemu}. The FlexKnot signal model can recover all tested signals with confidence and with reasonably low signal RMSE, and outperforms the Gaussian signal model when the true signal is not in the form of a Gaussian. Its ability to resolve narrow features that are not smooth, such as the flat bottom of a flattened Gaussian, however, has been shown to be weak.

In the presence of sinusoidal systematics as a proxy for unaccounted for systematics e.g. cable reflections, the FlexKnot signal model has the tendency to start fitting the systematics, as number of knots increases. This can be avoided by setting fewer number of knots. For the tested signals, which all have absorption level higher than 120 mK, the FlexKnot signal model can reliably recover the signal with confidence and reasonably low signal RMSE when the sinusoid as systematic has an amplitude lower than 50 mK. The deeper the absorption trough is, the less susceptible it is to the presence of systematics. The accuracy in finding the centre frequency, or the minimum point, of the true signal is also affected by sinusoidal systematics.

The optimal number of knots to recover the absorption history of the physical global signal is $N_{\mathrm{knot}}\sim 4-6$. When there are more than sufficient knots, the FlexKnot signal model would begin to fit structures that are unwanted such as the foreground chromaticity, yielding poorer fits. \citet{heimer23} arrived at a similar conclusion, showing that the evidence peaks at $N_{\mathrm{knot}}=6$ using EDGES low-band data despite the different implementations and interpolation methods. It is about the same for a signal in the form of a flattened Gaussian, as the FlexKnot signal model often struggles to fit the flat bottom and eventually recovers a signal alike to the other two types of signal that are smooth throughout. The optimal number of knots remains the same even in the presence of sinusoidal systematics, and adding more knots would result in it starting to fit other unwanted structures such as the systematics and the foregrounds.

Overall, we have shown that the FlexKnot signal model could be a better alternative to the existing models such as a Gaussian fit, having the potential to recover the absorption history of various different types that are smooth with reasonable confidence and accuracy in the likelihood analysis; it is capable of separating the signal as well as recovering the features given by the theoretical prediction. It also performs well in the framework of the REACH data analysis pipeline, where the foreground and the signal are jointly fitted.

\section*{Acknowledgements}
We thank Will Handley and Stefan Heimersheim for helpful discussions. ES is supported by Cambridge Trust and Taiwan Ministry of Education for their support. DA is supported by STFC and EdLA is supported by STFC Ernest Rutherford Fellowship. AF is supported by the Royal Society University Research Fellowship. 

\section*{Data Availability}



\bibliographystyle{mnras}
\bibliography{ref} 

\bsp	
\label{lastpage}
\end{document}